\documentclass[11pt]{article}
\usepackage[utf8]{inputenc}
\usepackage[margin=1.3in]{geometry}

\usepackage[T1]{fontenc}
\usepackage{kpfonts,baskervald}
\usepackage{authblk}
\usepackage{bbding}

\usepackage{fancyhdr}
\fancypagestyle{plain}{%
  \fancyhf{}%
  \fancyfoot[C]{--~\thepage~--}%
}
\usepackage{comment}
\usepackage{cite}

\usepackage{mathrsfs,mathtools}
\usepackage{amsfonts,amsmath,amssymb}
\usepackage{stmaryrd}
\usepackage{wasysym}

\usepackage{enumitem}
\usepackage{booktabs}
\usepackage{multirow}

\usepackage[all,cmtip]{xy}
\xyoption{all}

\usepackage{adjustbox}
\usepackage{tikz}
\usetikzlibrary{positioning,cd,decorations.markings}
\usepackage{pgfplots}
\usepackage[labelfont=bf]{caption}
\usepackage{blindtext}
\usepackage{dynkin-diagrams}

\numberwithin{equation}{section}

\usepackage{hyperref}

\definecolor{dgreen}{rgb}{0.09, 0.45, 0.27}
\definecolor{harvestgold}{rgb}{0.85, 0.57, 0.0}
\definecolor{airforceblue}{rgb}{0.36, 0.54, 0.66}
\hypersetup{
    colorlinks,
    citecolor=purple,
    filecolor=black,
    linkcolor=airforceblue,
    urlcolor=blue
}
\usepackage[nameinlink]{cleveref}
\crefname{thm}{lemma}{lemma}

\usepackage{amsthm}
\theoremstyle{definition}
\newtheorem{defn}{Definition}[section]
\newtheorem{example}{Example}[section]
\newtheorem{thm}{Theorem}[section]

\newtheorem{corollary}{Corollary}[section]
\newtheorem{lemma}[thm]{Lemma}

\theoremstyle{remark}
\newtheorem*{remark}{Remark}

\newcommand*\bbC{\mathbb{C}}
\newcommand*\bbD{\mathbb{D}}

\newcommand*\bbF{\mathbb{F}}

\newcommand*\bbH{\mathbb{H}}

\newcommand*\bbN{\mathbb{N}}

\newcommand*\bbP{\mathbb{P}}
\newcommand*\bbQ{\mathbb{Q}}
\newcommand*\bbR{\mathbb{R}}

\newcommand*\bbZ{\mathbb{Z}}

\newcommand*\diff{\mathop{}\!\mathrm{d}}

\newcommand*\I{\mathrm{i}}
\newcommand*\e{\mathrm{e}}

\newcommand*\cA{\mathcal{A}}
\newcommand*\cB{\mathcal{B}}
\newcommand*\cC{\mathcal{C}}
\newcommand*\cD{\mathcal{D}}
\newcommand*\cE{\mathcal{E}}
\newcommand*\cF{\mathcal{F}}

\newcommand*\cH{\mathcal{H}}

\newcommand*\cJ{\mathcal{J}}

\newcommand*\cL{\mathcal{L}}
\newcommand*\cM{\mathcal{M}}
\newcommand*\cN{\mathcal{N}}
\newcommand*\cO{\mathcal{O}}
\newcommand*\cP{\mathcal{P}}

\newcommand*\cS{\mathcal{S}}
\newcommand*\cT{\mathcal{T}}

\newcommand*\cV{\mathcal{V}}
\newcommand*\cW{\mathcal{W}}
\newcommand*\cX{\mathcal{X}}

\newcommand*\Frg{\mathfrak{g}}

\newcommand{\GL}{\mathrm{GL}}               
\newcommand*\U{\mathrm{U}}                  
\newcommand*\SU{\mathrm{SU}}                
\newcommand{\SL}{\mathrm{SL}}               
\newcommand{\PSL}{\mathrm{PSL}}             
\newcommand*\Sp{\mathrm{Sp}}                
\newcommand*\SO{\mathrm{SO}}                

\newcommand*\dih[1]{\mathrm{Dih}_{#1}}      
\newcommand{\lang}[1]{\prescript{L}{}{#1}}  

\newcommand{\jac}{\mathrm{Jac}}             
\newcommand*\ph[1]{\phantom{#1}}            
\newcommand*\Aut[1]{\mathrm{Aut}(#1)}       
\newcommand*\BPSL{\Lambda_\circ}            

\title{\bf{Global forms of $\mathcal{N}=4$ theories and non-minimal Seiberg-Witten solutions}}
\author[$\bullet$]{Robert Moscrop}
\affil[$\bullet$]{Center of Mathematical Sciences and Applications, Harvard University,

20 Garden Street, Cambridge, MA  02138,  USA}

\date{October 2025}
\pgfplotsset{compat=1.18} 

\begin{document}


\begin{titlepage}
\maketitle

\thispagestyle{empty}
\vspace{0.2cm}
\begin{abstract}
    \noindent To each four dimensional $\mathcal{N}\geq 2$ supersymmetric quantum field theory, one can associate an algebraic completely integrable (ACI) system that encodes the low energy dynamics of theory. In this paper we explicitly derive the appropriate ACI systems for the global forms of $\mathcal{N}=4$ super Yang-Mills (sYM) using isogenies of polarised abelian varieties. In doing so, we relate the complex moduli of the resulting varieties to the exactly marginal coupling of the theory, thus allowing us to probe the $S$-duality groups of the global forms. Finally, we comment on whether the resulting varieties are the Jacobians of a minimal genus Riemann surface, coming to the conclusion that many global forms of $\mathcal{N}=4$ sYM do not admit a minimal genus Seiberg-Witten curve that correctly reproduces the global form. 
\end{abstract}
\vfill

\begin{flushright}
{}\noindent $\underline{\hspace{4.5cm}}$\\
$^\bullet$\footnotesize{{\tt \href{mailto:robert@cmsa.fas.harvard.edu}{robert@cmsa.fas.harvard.edu}}}
\end{flushright}
\end{titlepage}


\newpage
\setcounter{page}{1}
{\hypersetup{linkcolor=black}
\tableofcontents
}


\section{Introduction}

Given a complete set of local operators, there may be several quantum field theories (QFTs) which realise them. In order to distinguish between these various QFTs, one has to specify the spectrum of extended objects of the theory \cite{Gaiotto:2010be}. In the case of $4d$ gauge theory, specifying the extended operators is equivalent to specifying the global form of the gauge group along with some additional discrete data \cite{Aharony:2013hda}. We refer to a theory obtained in this manner as a \emph{global form} or \emph{absolute} version of the relative theory specified by the local dynamics. With this in mind, it is natural to ask how the various structures associated to a theory reflects this choice in global form, if at all. Of particular interest to us, in the case of $4d$ $\cN=2$ theories, how does the geometry of the moduli space depend on this?

Recall that the moduli spaces of $4d$ $\cN=2$ supersymmetric quantum field theories (SQFTs) contain a distinguished $r$-dimensional submanifold $\cC$ called the Coulomb branch (CB) whose vacua correspond to a low energy $\U(1)^r$ gauge theory. At any point of the Coulomb branch, one has access to several objects of physical interest: the lattice of electromagnetic charges of BPS states $\Lambda$, the matrix of effective couplings $\tau$ and the Dirac pairing $J:\Lambda\times\Lambda\rightarrow \bbZ$ measuring the mutual non-locality of states. Writing $J$ in the form
\begin{gather}
    J=\left[\begin{array}{cc}
        \ph{-}\mathbf{0} & D  \\
         -D &  \mathbf{0}
    \end{array}\right],\quad D=\mathrm{diag}(d_1,\ldots,d_r). 
\end{gather}
allows us to define an $r$-dimensional abelian variety with period matrix $\Pi=[D,\tau]$ at any smooth point of the CB. This results in a fibration of polarised abelian varieties over $\cC$ which, when supplemented with a particular holomorphic 2-form, defines an algebraic completely integrable (ACI) system encoding the low energy dynamics of the theory \cite{Donagi:1995cf,Donagi95,vanhaecke96}. This construction neglects any possible extended objects of the theory and is therefore insensitive to the global form of theory. However, one can define an analogous system for the absolute versions of the theory by refining the BPS charge lattice to include additional charges corresponding to probe lines \cite{Gaiotto:2010be,Argyres:2022kon}. At the level of the abelian fibres, this produces a new abelian variety $A_{\mathrm{abs}}$ that is related to the relative fibre by a map known as an \emph{isogeny}. As a result, the period matrix of the new variety is given by
\begin{gather}\label{eq:change}
    \Pi_{\mathrm{abs}} = G \Pi R,
\end{gather}
where $G\in \GL_r(\bbC)$ and $R\in\mathrm{Mat}(2r\times 2r,\bbZ)$ encode how the BPS charge lattice embeds into its refinement. Crucially, the refinements must result in a principal polarisation on $A_{\mathrm{abs}}$, so the possible refinements are completely determined by $J$ and $\Lambda$. In this paper, we will rephrase the results of \cite{Gaiotto:2010be,Argyres:2022kon} in terms of abelian varieties and isogenies, as above, and explicitly perform this procedure in the case of $\cN=4$ super Yang-Mills (sYM). Noting that \cref{eq:change} describes the structure of the absolute varieties in terms of the exactly marginal coupling of $\cN=4$ sYM, we then check our results by calculating the $S$-duality orbits of the global forms purely from the point of view of the fibres. Note that the procedure of using isogenies to construct absolute geometries has been studied extensively in rank-1 \cite{Closset:2023pmc} where the fibres of the ACI system are elliptic curves. Our work therefore gives a partial generalisation of this and provides many higher rank examples.

As the procedure for obtaining the ACI systems for absolute versions of a theory result in a principally polarised abelian variety, it is interesting to ask if it is the Jacobian of a Riemann surface. Our motivation for asking this question stems from the ongoing program to classify the possible $4d$ $\cN=2$ superconformal field theories (SCFTs) (see \cite{Argyres:2020nrr,Martone:2021ixp} for recent reviews). In recent years, much progress has been made by probing the possible Seiberg-Witten curves \cite{Seiberg:1994aj,Seiberg:1994rs} of low rank theories, resulting in a complete classification of rank-1 SCFTs and an increasingly well explored landscape of rank-2 theories. In these cases the corresponding ACI system is constructed using the Jacobians of the genus $r$ Riemann surfaces together with the differential of the Seiberg-Witten 1-form. However, as not all principally polarised abelian varieties of dimension greater than $3$ are Jacobians, the existence of a minimal genus Seiberg-Witten curve can fail when the rank exceeds $3$. By remarking that the Hurwitz automorphism theorem severely restricts the possible automorphism groups of Jacobians of smooth curves, we will show that various global forms of $\cN=4$ sYM theories do not admit a minimal genus Seiberg-Witten solution if one wishes to be able to distinguish between the various global forms of the theory. These examples include $\cN=4$ sYM with gauge groups $F_4$ and $E_8$ which only admit one global form. As such, these examples show the limitations of classifying SCFT geometries via minimal Seiberg-Witten solutions. 

\medskip
\noindent {\bf Outline.} In section 2 we present a pedagogical overview of abelian varieties and outline the Schottky problem. Appendix A also provides further material on the moduli spaces of abelian varieties and fixes our conventions for isomorphisms of such objects. Section 3 recasts the refinement process of the BPS charge lattice in the language of isogenies and polarisations of abelian varieties, before moving onto giving several examples of this isogeny process. In particular, we explicitly find all smooth fibres of the ACI systems for $\SU_N$ and $B_2$ $\cN=4$ sYM. We then show that our results allow us to easily extract the $S$-duality orbits of the different global forms of the theory through isomorphisms of PPAVs. Finally, in section 4 we comment on whether the abelian varieties found in section 3 can be understood as the Jacobian of a Riemann surface of minimal genus, coming to the conclusion that many absolute $\cN=4$ theories do not possess such a Seiberg-Witten curve which reproduces the physical ACI system via Jacobians.

\medskip
\noindent {\it Note.} During final stage of preparation for this paper, I was made aware of the works \cite{SKLines, Cecotti:2025lns} whose subject overlaps with our work (particularly section 3). Our results agree (almost) unanimously, though the methodologies and aims of the works differ. For the sake of clarity, we detail the differences and overlap here.
\begin{itemize}
    \item The approach of \cite{SKLines} leverages knowledge of the integral representation theory of Weyl groups to understand the special K\"ahler and $S$-duality structures of absolute $\cN=4$ theories. The only discrepancy in our results concerns the $S$-duality group when $\Frg=A_1$ and $\Frg=B_2$. This stems from the fact we define the action on the marginal coupling of the theory by studying the relative ACI system and then tracking how it breaks to the absolute theories. As the analysis of \cite{SKLines} is agnostic to the relative geometry, their absolute $S$-duality groups instead include all self-identifications regardless of if they stem from the relative geometry. In other words, the transformations considered in \cite{SKLines} act on the complex modulus of the absolute fibre and do not take into account how this is related to the coupling defined by relative geometry, thus leading to additional identifications.
    \item The aim of \cite{Cecotti:2025lns} is to explain the physical meaning of why some special K\"ahler geometries do not admit crepant resolutions, using the class of $*$-isotrivial geometries to explore this. Theories with $\cN=4$ supersymmetry are contained within this class and are a prominent example within the paper. The construction of absolute fibres via isogenies from certain `root' varieties is explored and used as a technical tool to eliminate torsion from the homology group of the central (most singular) fibre after normalisation. Our results regarding the absolute geometries of $\cN=4$ theories agree. 
\end{itemize}
While these works overlap with the material presented in section 3, I believe this work still provides useful insight into the geometry of $\cN=4$ theories and gives several explicit examples of the procedure to obtain such geometries. Additionally, I thank the authors of \cite{SKLines} for coordinating the submission of our articles to the arXiv.

\section{Preliminaries}

\subsection{Complex tori and isogenies}

Let $V=\bbC^g$ and $\Lambda\subset V$ be a lattice of rank $2g$ inside $V$. The quotient space $X=V/\Lambda$ is a compact abelian Lie group called a \emph{complex torus}. A useful way of encoding a complex torus is as follows. If $V$ has a basis $\{v_1,\ldots,v_g\}$ and $\Lambda$ has basis $\{\lambda_1,\ldots,\lambda_{2g}\}$, the matrix $\Pi$ defined by the relation
\begin{gather}
    \lambda_j = \sum_{i=1}^{g} \Pi_{ji}v_i,
\end{gather}
is called a \emph{period matrix} and it completely determines the complex torus as $X=\bbC^g/\Pi \bbZ^{2g}$. Clearly, the period matrix depends on the choice of bases for $V$ and $\Lambda$, so it is often convenient to choose the bases such that $\Pi=[\mathrm{id}_g,Z]$ where $Z$ is a $(g\times g)$-dimensional matrix with $\det(\mathrm{Im}\,Z)\neq 0$. In fact, there always exists bases for $V$ and $\Lambda$ that put the period matrix in this form \cite[prop. 1.1]{BL_tori}.

Now let $X'=V'/\Lambda'$ be another complex torus of dimension $g'$ with a period matrix $\Pi'$. Any homomorphism $f:X\rightarrow X'$ can be lifted to a unique linear map $F$ between the covering spaces $V$ and $V'$ satisfying $F(\Lambda)\subset \Lambda'$ \cite{BL92}. This defines a homomorphism
\begin{gather}
    \rho_a:\mathrm{Hom}(X,X')\rightarrow \mathrm{Hom}_\bbC(V,V'), \quad \rho_a(f)=F,
\end{gather}
called the \emph{analytic representation} of $\mathrm{Hom}(X,X')$. Similarly, by restricting $F$ to $\Lambda$, we obtain another homomorphism
\begin{gather}
    \rho_r:\mathrm{Hom}(X,X')\rightarrow \mathrm{Hom}_\bbZ(\Lambda,\Lambda'), \quad \rho_r(f)=F|_\Lambda,
\end{gather}
which is known as the \emph{rational representation}. By noting that the period matrices of $X$ and $X'$ define embeddings of the lattices into the covering spaces, we see that the analytic and rational representations of $f$ fit into the commutative diagram
\begin{equation}
    \begin{tikzcd}
        \bbZ^{2g} \arrow[hook,"\Pi"]{r} \arrow["\rho_r(f)" left]{d}& \bbC^g \arrow["\rho_a(f)"]{d}\\
        \bbZ^{2g'} \arrow[hook,"\Pi'"]{r} & \bbC^{g'}
    \end{tikzcd}
\end{equation}
Thus giving us the relation
\begin{gather}
    \rho_a(f)\Pi = \Pi'\rho_r(f).
\end{gather}
Conversely, any set of matrices $A\in \mathrm{Mat}(g'\times g,\bbC)$ and $R\in \mathrm{Mat}(2g'\times2g,\bbZ)$ satisfying $A\Pi=\Pi'R$ define a homomorphism between $X$ and $X'$.

A particularly important class of maps is that of \emph{isogenies}. These are surjective morphisms between complex tori with finite kernel, so $V\cong V'$ necessarily. The prototypical example of an isogeny is the multiplication map $n_X$ given by
\begin{gather}
    n_X : X\rightarrow X,\quad x\mapsto nx,\quad n\in\bbN.
\end{gather}
The kernel, denoted by $X[n]$, is called the set of $n$-torsion points and has the structure $X[n]=\bbZ_{n}^{2g}$. Crucially, every isogeny has an `inverse up to torsion'. That is, given an isogeny $f:X\rightarrow X'$ with kernel of exponent $q$, there exists an isogeny $g:X'\rightarrow X$ such that $gf=q_X$ and $fg=q_{X'}$ \cite[prop. 1.2.6]{BL92}. As such, isogenies define an equivalence relation on the set of complex tori and if two tori $X$ and $Y$ are fall into the same class, we call them \emph{isogenous}. This will be denoted as $X\simeq Y$.

\subsection{Polarisations}
Note that, in general, a complex torus is not a projective variety. If a complex torus $X$ has the structure of a projective variety it is said to be an \emph{abelian variety}. By Chow's theorem, it suffices to exhibit an embedding $\varphi: X \hookrightarrow \bbP^N$ to establish this property. In order to do this, one often endows $X$ with a \emph{polarisation}, which we now review.

To form a map into $\bbP^N$, we require $(N+1)$-many independent globally defined functions $\{s_i:X\rightarrow \bbC:i=0,\ldots,N\}$ so we can define
\begin{gather}\label{eq:embed}
    \varphi: x \mapsto [s_0(x):s_1(x):\cdots:s_{N-1}(x):s_N(x)]\in \bbP^N.
\end{gather}
We can think of these functions $s_i$ as belonging to the space of global sections $H^0(\cL)$ of some line bundle $\cL$. Of course, not all line bundles will have the property that \cref{eq:embed} is an embedding, so we call line bundles \emph{very ample} if \cref{eq:embed} embeds $X$ into $\bbP^N$. If a bundle $\cL$ is not very ample, but a tensor power of it is, then one calls $\cL$ merely \emph{ample}.

Recall that the group of holomorphic line bundles $\mathrm{Pic}(X)$ on a space $X$ is given by $H^1(X,\cO_X^*)$ where $\cO_X^*$ is the sheaf of nowhere vanishing functions on $X$. Using the long exact sequence in cohomology associated to
\begin{gather}
    1\rightarrow \bbZ \rightarrow \cO_X\rightarrow \cO_X^*\rightarrow 1,
\end{gather}
we can introduce the first Chern class as the connecting morphism between
\begin{gather}
    c_1 : H^1(X,\cO^*_X)\rightarrow H^2(X,\bbZ).
\end{gather}
The image of $c_1$ is called the \emph{N\'eron-Severi group} $\mathbf{NS}(X)$ and the kernel is denoted by $\mathrm{Pic}^0(X)$. The Kodaira embedding theorem states that a line bundle $\cL$ on $X=V/\Lambda$ is ample if and only if its first Chern class can be represented by a positive definite Hermitian form $H$ on $V$ satisfying $\mathrm{Im}\,H(\Lambda,\Lambda)\subset \bbZ$. Noting that $\mathrm{Im}\,H$ is an antisymmetric form on $\Lambda$, we can regard the N\'eron-Severi group as either the set Hermitian forms described above, or the set of alternating forms $E:V\times V\rightarrow\bbR$ satisfying $E(\Lambda,\Lambda)\subseteq \bbZ$ and $E(\I u,\I v)=E(u,v)$ for any $u,v\in V$. This leads us the definition of a polarisation.

\begin{defn}\label{def:pol}
    Let $X=V/\Lambda$ be a complex torus. A \emph{polarisation} on $X$ is the class $H$ of an ample line bundle $\cL$ in the N\'eron-Severi group $\mathbf{NS}(X)$ representing $c_1(\cL)$. The pair $(X,\cL)$ is the called a \emph{polarised abelian variety}.
\end{defn}

\noindent Note that there always exists a basis of $\Lambda$ such that $E=\mathrm{Im}\, H$ takes the form
\begin{gather}\label{eq:symplectic}
    E = \left[\begin{array}{cc}
        \ph{-}\mathbf{0} & D \\  
         -D& \mathbf{0}
    \end{array}\right],\quad D=\mathrm{diag}(d_1,\ldots,d_g),
\end{gather}
where $d_i|d_{i+1}$. Such a basis is called a \emph{symplectic basis} and the set $\{d_i\}$ are called the \emph{invariant factors} of $\cL$. In particular, if a polarisation's invariant factors are all $1$, we say that the polarisation is \emph{principal}. Furthermore, after specifying a symplectic basis, a basis of $V$ can be chosen such that the period matrix takes the form $\Pi=[D,Z]$ with $Z$ symmetric and $\mathrm{Im}\, Z$ positive definite. When in a symplectic basis, we refer to $Z$ as the complex modulus of the variety.

Before moving on, let us provide a second characterisation of a polarisation which will be useful later. As a consequence of the Appell-Humbert theorem, the dual torus $\hat{X}=V^*/\Lambda^*$ can be identified with $\mathrm{Pic}^0(X)$ \cite[prop. 2.4.1]{BL92}. Given any line bundle $\cL$ on $X$, we can define the map
\begin{gather}
    \phi_{\cL} : X\rightarrow \hat{X},\quad \phi_{\cL}(x) = t_x^* \cL\otimes \cL^{-1}\in\mathrm{Pic}^0(X),
\end{gather}
where $t_x$ denotes translation by $x\in X$. This map is a homomorphism of complex tori and an isogeny if and only if $H=c_1(\cL)$ is non-degenerate. In fact, the morphism $\phi_\cL$ only depends on the first Chern class of $\cL$ and its analytic representation is given by $v\mapsto H(v,\cdot)$. 

Now consider the kernel $K(\cL)$ of $\phi_\cL$. This measures the difference between the dual lattice and $\Lambda$ under the rational representation of the polarisation. In other words, we can characterise $K(\cL)$ as
\begin{gather}
    K(\cL) = \Lambda(\cL)/\Lambda,\quad \Lambda(\cL) = \{v\in V:E(v,\Lambda)\subseteq \bbZ\}.
\end{gather}
By taking a symplectic basis of $\Lambda$, as in \cref{eq:symplectic}, we see that
\begin{gather}\label{eq:kernel}
    K(\cL) = (\bbZ^g/D\bbZ^g)\oplus (\bbZ^g/D\bbZ^g).
\end{gather}
Importantly, $K(\cL)$ is endowed with a natural pairing $\e^\cL$ defined via $\Lambda(\cL)$ as
\begin{gather}
    \e^\cL(v_1,v_2) = \exp(-2\pi\I E(v_1,v_2)).
\end{gather}
As $E$ is integral on $\Lambda$, it is clear that this descends to a well-defined pairing on $K(\cL)$. In particular, each $\bbZ^g/(D\bbZ^g)$ factor in \cref{eq:kernel} is isotropic with respect to $\e^\cL$ and dual to each other, thus making the pairing perfect.

Suppose two complex tori $X=V/\Lambda$ and $X'=V'/\Lambda'$ are isogenous $f:X\rightarrow X'$ and that $X'$ possesses a polarisation $\cL$. As the rational representation of $f$ maps $\Lambda$ into $\Lambda'$, we can restrict $E= \mathrm{Im}\,c_1(\cL)$ to this sublattice to obtain an integral non-degenerate pairing on $\Lambda$. Therefore, the pullback $f^*\cL$ provides a polarisation on $X$. In the opposite direction, supposing $X$ has a polarisation $\cN$, we can ask if there exists a polarisation $\cL$ on $X'$ such that $\cN=f^*\cL$. The existence of $\cL$ is not always guaranteed and the condition for existence can be summarised as follows \cite[cor. 6.3.5]{BL92}.

\begin{lemma}\label{lem:isotropic}
    Let $\cN$ be a polarisation on $X$. Given an isogeny $f:X\rightarrow X'$, there exists a polarisation $\cL$ on $X'$ satisfying $\cN=f^*\cL$ if and only if $\ker f$ is an isotropic subgroup of $K(\cN)$ with respect to the pairing $\e^\cN$.
\end{lemma}

\noindent In terms of the lattices, this condition ensures that the antisymmetric pairing $E$ on $\Lambda$ extends to an integral pairing on $F^{-1}(\Lambda')$ where $F=\rho_a(f)$.

\subsection{Jacobians and the Schottky problem}\label{sec:jac}

Given a smooth compact Riemann surface $C$ of genus $g$, there is a canonical way to construct an abelian variety known as the \emph{Jacobian variety} $\jac(C)$. First of all, note that any class $[\gamma]$ in $H_1(C,\bbZ)$ defines a map
\begin{gather}
    [\gamma]:H^0(\Omega_C)\rightarrow\bbC,\quad \omega\mapsto\int_\gamma \omega,
\end{gather}
where $\Omega_C$ is the sheaf of holomorphic forms on $C$. It can be shown that this gives an embedding of $H_1(C,\bbZ)$ into $H^0(\Omega_C)^*$, allowing us to define
\begin{gather}
    \jac(C)=H^0(\Omega_C)^*/H_1(C,\bbZ).
\end{gather}
As $H^0(\Omega_C)\cong \bbC^g$ and $H_1(C,\bbZ)$ is a rank $2g$ lattice, $\jac(C)$ has the structure of a complex torus. Furthermore, the intersection pairing on $H_1(C,\bbZ)$ provides a principal polarisation $\Theta_C$ on $\jac(C)$ called the \emph{canonical polarisation}. The importance of this polarisation is highlighted by the following theorem of Torelli.

\begin{thm}
    {\bf (Torelli \cite{tor1913}).} Any curve $C$ is uniquely determined by its Jacobian $\jac(C)$ together with its canonical polarisation $\Theta_C$.
\end{thm}

\noindent By relaxing the canonical polarisation requirement, one finds that the theorem fails. Indeed, one can find many examples of non-isomorphic curves whose Jacobians are isomorphic as complex tori only \cite{howe98,howe03}. Further note that it can be shown that $\Theta_C$ is irreducible \cite{HM63}, so $\jac(C)$ is always an indecomposable abelian variety (but may still split as a complex torus).

Now consider the map $\cT:C\mapsto(\jac(C),\Theta_C)$ as mapping between the moduli space of compact genus $g$ Riemann surfaces $\cM_g$ and the moduli space of indecomposable $g$-dimensional principally polarised abelian varieties $\cA_g^{\mathrm{in}}$. Torelli's theorem tells us that this map is injective and it can be shown that it is an immersion \cite{Oort80}, so $\cT$ embeds $\cM_g$ into $\cA_g^{\mathrm{in}}$. It is therefore natural to ask for a characterisation of the (open) \emph{Torelli locus} $\cT_g=\cT(\cM_g)$.\footnote{Note that some authors refer to the Zariski closure of $\cT_g$ as the Torelli locus. The additional varieties in $\overline{\cT_g}$ constitute products of lower dimensional Jacobians. As this distinction will important to us later, we will make it clear when we are dealing with the closure of $\cT_g$.} This is known as the Schottky problem.

Before going into further details, recall that for $g>1$ 
\begin{gather}
    \dim\cM_g=3g-3,\quad\quad \dim\cA_g=\tfrac{1}{2}g(g+1),
\end{gather}
from which we see that the dimension of these spaces match for $g=2$ and $g=3$, while $\cT_g$ is of codimension $\tfrac{1}{2}(g-2)(g-3)$ for $g>3$. This gives us an indication that the Schottky problem for $g>3$ is distinctly different to the $g\leq 3$ case. Let us now summarise some more precise results about the Schottky problem for low dimensions.
\begin{itemize}
    \item For $g\leq 3$, we know that the dimensions of $\cA_g$ and $\cM_g$ match, but we can make an even sharper statement. Due to a theorem of Ueno and Oort \cite{OortUeno73}, every indecomposable principally polarised abelian variety of dimension 3 or less is isomorphic to the Jacobian of a smooth Riemann surface. As such, $\cT_g = \cA_g^{\mathrm{in}}$ for $g\leq 3$.
    \item In four dimensions, we see that $\cT_4$ is of codimension 1 inside $\cA_4$. Schottky originally proposed that a particular modular form vanished when evaluated on the Zariski closure of $\cT_4$  \cite{Schottky1888}.  Igusa later proved this was the case, further showing that the vanishing locus of this modular form actually defines $\overline{\cT_4}$ as a hypersurface in $\cA_4$ \cite{igusa81}.
    \item For $g\geq 5$, the Schottky problem is still open and various approaches beyond those used in $g=4$ have been developed to study it. We refer readers to \cite{MR2931868} for a detailed, yet approachable, review of the status of the problem.
\end{itemize}

While a full solution to the Schottky problem is unknown, various facts about $\cT_g$ for any value of $g$ are known. In particular, due to Torelli's theorem, the automorphisms of a Riemann surface are closely related the automorphisms of its Jacobian via the following corollary.

\begin{corollary}\label{cor:autojac}
    Let $C$ be a smooth Riemann surface of genus $g>1$. Then the group of automorphisms is given by
    \begin{gather}
        \Aut{C} = \begin{cases}
            \Aut{\jac(C),\Theta_C} & \text{if }C \text{ is hyperelliptic},\\
            \Aut{\jac(C),\Theta_C}/\langle -\mathrm{id}\rangle &\text{otherwise.}
        \end{cases}
    \end{gather}
\end{corollary}

\begin{figure} 
    \centering
    \includegraphics[scale=0.28,decodearray={0 1 0.2 1 0.2 1}]{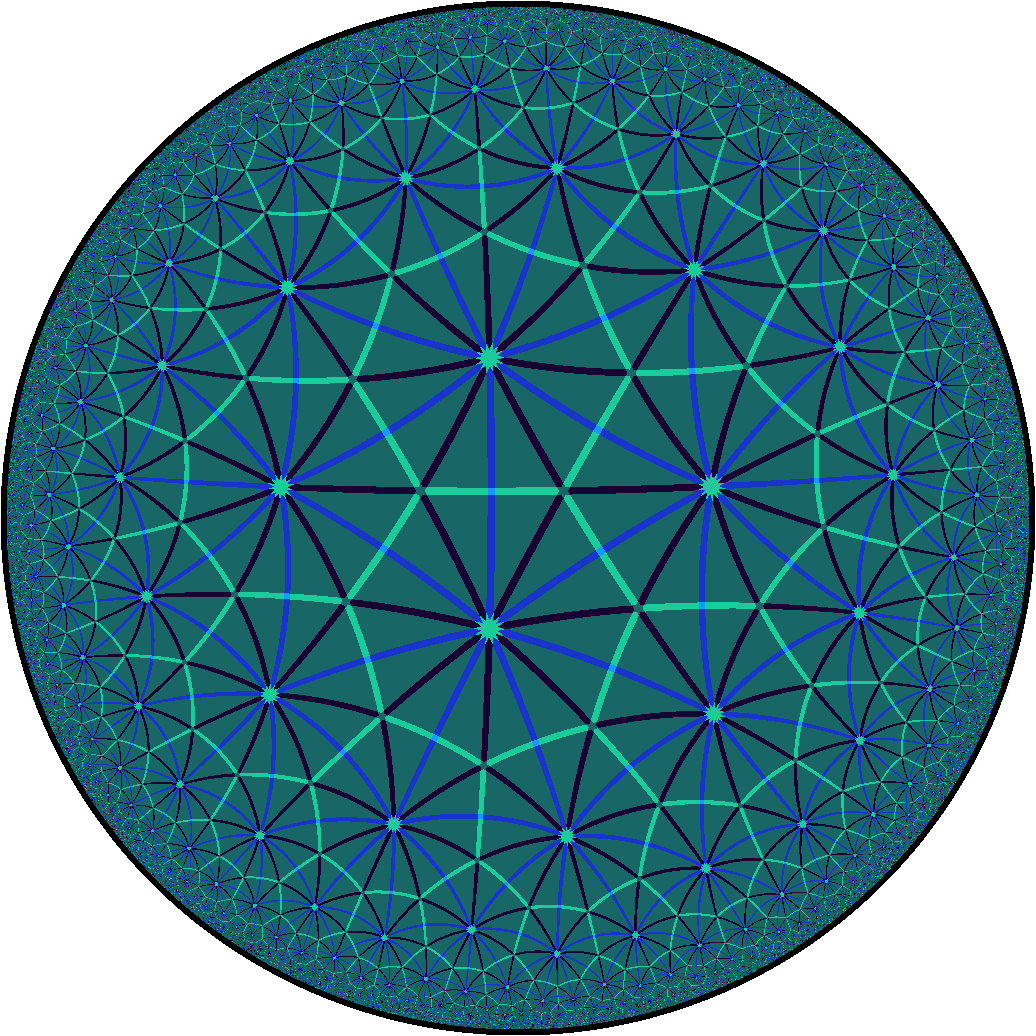}\quad\quad
    \includegraphics[scale=0.28,decodearray={0 1 0.2 1 0.2 1}]{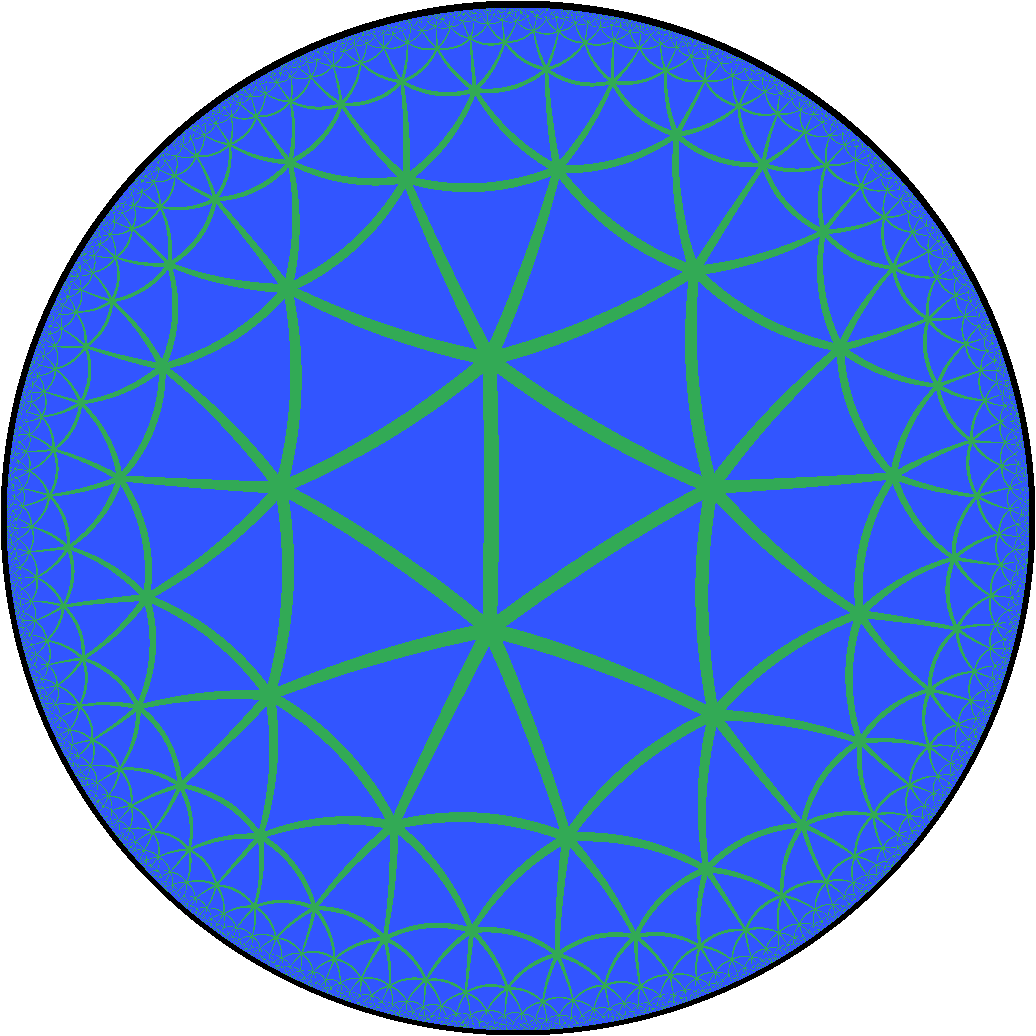}
    \caption{\emph{Left}: The tiling of the Poincar\'e disk by the Schwarz triangle with internal angles $\pi/2$, $\pi/3$ and $\pi/7$. \emph{Right}: The order-$7$ triangular tiling of the Poincar\'e disk. Every Hurwitz surface admits a triangulation which is a quotient of this tiling. This image was generated with the Python code available at \cite{sch_py}.}
    \label{fig:ord7}
\end{figure}

\noindent As such, knowledge of the automorphism groups of Riemann surfaces can be used to constrain the Torelli locus. These groups are highly constrained due to the Hurwitz automorphism theorem, which we briefly review. 

In order to maximise the automorphism group of a Riemann surface, it is useful to instead try to minimise the area of the fundamental domain of the action of the group. By virtue of Riemann uniformisation, a tiling of fundamental domains for the automorphism group of a genus $g>1$ Riemann surface can be lifted to a polygonal tiling of the Poincar\'e disk. Minimising the area of each tile maximises the automorphism group after closing the tiling and returning to the original Riemann surface. This can be achieved by tiling the disk by Schwarz triangles with interior angles $\pi/2$, $\pi/3$ and $\pi/7$, but in order to exclude orientation reversing automorphisms, one must glue copies of these triangles together, resulting in the tiling on the right of \cref{fig:fund_domain}. This is captured by the following theorem.

\begin{thm}
    {\bf (Hurwitz \cite{hur1892}).} Let $C$ be a smooth compact genus $g\geq 2$ Riemann surface. Then
    \begin{gather}
        |\Aut{C}| \leq 84(g-1).
    \end{gather}
    A surface which saturates this bound is called a {\it Hurwitz surface} the corresponding automorphism group a {\it Hurwitz group}.
\end{thm}
\noindent The covering space point of view tells us that the Hurwitz groups are finite quotients of symmetry group of the order-7 triangular tiling. As such, Hurwitz groups can be obtained as quotients $(2,3,7)$-triangle group
\begin{gather}
    \triangle(2,3,7)= \{a,b: a^2=b^3=(ab)^7=1\},
\end{gather}
examples of which include $\PSL_2(\bbF_7)$ and $\PSL_2(\bbF_8)$. It is worth noting that Hurwitz surfaces do not occur at every genus. Indeed, no curve saturates the bound for $g=2$, but the Klein quartic does in genus $3$.

Returning to the moduli spaces, denote the subset of $\cA_g^{\mathrm{in}}$ of principally polarised varieties with automorphism groups larger than $168(g-1)$ by $\cV_g$. By using \cref{cor:autojac} with the Hurwitz automorphism theorem, we see that any variety belonging to $\cV_g$ cannot be the Jacobian of a smooth genus $g$ Riemann surface. Therefore, in terms of the Torelli locus, we have $\cT_g\cap \cV_g = \varnothing$.

\section{ACI systems for absolute \texorpdfstring{$\cN=4$}{N=4} theories}

\subsection{Refinements and isogenies}\label{sec:refine}

Within the moduli space of a given $\cN=2$ SQFT there is a distinguished submanifold $\cC$ called the \emph{Coulomb branch}. The vacua forming $\cC$ are characterised by a low energy $\mathrm{U}(1)^r$ gauge theory with a non-zero mass gap. We can write the bosonic part of the effective action as
\begin{gather}
    L_\cC = \mathrm{Im}\left[\tau_{ij}(a)(\partial a^i\cdot \partial\bar{a}^j)+\cF_i\cdot\cF_j\right],
\end{gather}
where $\cF_i=\tfrac{1}{2}(F_i+\star F_i)$ is the self-dual field strength of the $i$-th $\mathrm{U}(1)$ factor, $\{a^i\}$ are the complex scalars belonging to the vector multiplets called the \emph{special coordinates} and $\tau_{ij}(a)$ the matrix of effective couplings. Using the field strengths, one can then define electric and magnetic charges as $e_i=\oint_{S^2}\star F_i$ and $g_i=(2\pi)^{-1}\oint_{S^2} F_i$. 

Note that the Coulomb branch can display metric singularities along a locus $\Delta\subset\cC$. Vacua belonging to $\Delta$ contain additional massless states called \emph{BPS states}. By probing all such states of the theory, one can define the \emph{BPS charge lattice} $\BPSL\cong \bbZ^{2r}$ to be the lattice generated by the charges of the BPS states of the theory. This lattice, by Dirac quantisation, is equipped with an integral non-degenerate alternating form $\cJ:\BPSL\times\BPSL \rightarrow \bbZ$ which gives $\BPSL$ the structure of a symplectic lattice.

The physical data $(\BPSL,\tau(a),\cJ)$ allows us to define an abelian variety $A_{\tau(a)}$ over each smooth point $a$ of the Coulomb branch --- the alternating form $\cJ$ gives $A_{\tau(a)}$ a natural polarisation while $\tau(a)$ describes the complex modulus of $A_{\tau(a)}$. In particular, if $\cJ$ has invariant factors $D=\mathrm{diag}(d_1,\ldots,d_r)$, the period matrix of $A_{\tau(a)}$ is given by $\Pi=[D, \tau(a)]$. However, If we approach a singular point $u\in\Delta$, $A_{\tau(a)}$ similarly develops singularities and fails to be an abelian variety. Overall, this gives us a fibration $\pi: \cX\rightarrow \cC$ over the Coulomb branch with generic fibres polarised abelian varieties. Furthermore, this fibration can be endowed with a holomorphic $2$-form $\omega\in H^{(2,0)}(\cX)$ whose restriction to the fibres vanishes --- that is, the fibration is Lagrangian with respect to $\omega$.\footnote{The 2-form $\Omega$ also has physical importance --- integrals over 1-cycles in $A_{\tau(a)}$ give the differentials of the special coordinates and their duals.} This is precisely the structure of an \emph{algebraic completely integrable (ACI) system} \cite{Donagi:1995cf,Donagi95,vanhaecke96}.

While the ACI system above encodes the low energy dynamics of the theory, there is a subtlety that arises when one considers the possible line operators of a theory. For any charge, regardless of whether it belongs to $\BPSL$ or not, one can define a probe line operator by specifying particular boundary conditions for fields as they approach the support of the line operator \cite{Kapustin:2005py}. The electromagnetic charges carried by these line operators are interpreted as the charges of the theory's $1$-form symmetry and are restricted by Dirac quantisation which imposes that they must have integral Dirac pairing with any charge in $\BPSL$. As such, the charges of possible line operators belong to the lattice dual to $\BPSL$ under the pairing $\cJ$. We therefore interpret the kernel of the polarisation isogeny
\begin{gather}
    K(\cJ) = \BPSL(\cJ)/\BPSL, \quad \BPSL(\cJ) =\{v\in V:\cJ(v,\BPSL)\subseteq \bbZ\},
\end{gather}
as the group of possible line operator charges up to screening by BPS states. In particular, this is the defect group of \cite{DelZotto:2015isa,Albertini:2020mdx}. As pointed out in \cite{Gaiotto:2010be}, when $K(\cJ)$ is non-trivial, $\BPSL$ does not fully define the theory and one should specify a spectrum of maximally commuting line operators to obtain a fully defined theory. Instead, $\BPSL$ specifies a \emph{relative} theory in the language of \cite{Freed:2012bs} and including an appropriate spectrum of lines gives us an \emph{absolute} theory. By specifying this spectrum of lines, the BPS charge lattice is refined to include these additional charges in such a way that the resulting polarisation is principal \cite{Argyres:2022kon}.

At the level of the ACI system, the refinement of the BPS charge lattice leads to an isogeny $f:A_{\tau(a)}\rightarrow X_{\tau(a)}$ with the property that $X_{\tau(a)}$ possesses a principal polarisation $\cP$ such that the pullback of $\cP$ along $f$ gives the natural polarisation $\cJ$ on $A_{\tau(a)}$. In terms of the associated isogenies, this is summarised in the commutative diagram:
\begin{gather}\label{eq:comm}
\begin{aligned}
    \xymatrix@R=2.5pc@C=2.5pc{
    \hat{X}  & \hat{Y} \ar[l]_{\hat{f}} \\
    X \ar[u]^{\phi_\cJ} \ar[r]^{f} & Y \ar[u]_{\phi_{\cP}}
   }
\end{aligned}
\end{gather}
Here $\hat{f}$ denotes the isogeny dual to $f$ whose rational representation is $\rho_r(\hat{f})=\rho_r(f)^{T}$. In particular, the degree of $\hat{f}$ is equal to the degree of $f$. As all maps in \cref{eq:comm} are surjective, we get
\begin{gather}
    \deg\phi_{\cJ}=(\deg f)^2\cdot\deg\phi_{\cP}.
\end{gather}
Since the natural polarisation on $\BPSL$ encodes the defect group $K(\cJ)$, we see that in order for $\cP$ to be principal, we must have
\begin{gather}
    \deg f = \prod_i d_i.
\end{gather}
Furthermore, \cref{lem:isotropic} further restricts $\ker f$ to be an isotropic subgroup of the defect group $\bbD=K(\cJ)$. Physically, this condition appears in the geometric engineering viewpoint of higher form symmetries where it corresponds to picking out a maximally commuting subgroup of the Freed-Segal-Moore fluxes \cite{Freed:2006yc} as discussed in \cite{Albertini:2020mdx}. In summary, absolute ACI systems are obtained from the relative system by performing fibre-wise isogenies whose kernels are maximally isotropic subgroups of the defect group that specify the charges of probe line operators.

\subsection{Fibres for \texorpdfstring{$\cN=4$}{N=4} theories}

\subsubsection{The relative fibres}\label{sec:relative}

Let us now restrict to $\cN=4$ sYM with gauge algebra $\Frg$. It can be shown, by going to the weak coupling limit, that the electric charges of the theory in a Coulomb vacuum span the root lattice $\Gamma_r$ of $\Frg$, while the magnetic charges span the coroot lattice $\Gamma_r^\vee$, so the full set of BPS charges is given by $\Gamma_r\oplus\Gamma_r^\vee$\cite{Kapustin:2005py}. Additionally, there is an exactly marginal coupling $\tau=4\pi\I /g_{\mathrm{YM}}^{2}+\theta/2\pi$ which enters into the coupling matrix $\tau(a)$ and, therefore, the period matrix of the corresponding abelian fibre. To make this explicit, we define the BPS charge lattice $\BPSL$ to be 
\begin{gather}
\BPSL\doteq\Gamma_r+\tau\Gamma_r^\vee\cong\Gamma_r\oplus\Gamma_r^\vee\,\quad \tau\in\bbH_1.
\end{gather}
Due to this simple form, the corresponding complex tori $A_\tau=\bbC^n/\BPSL$ are particularly easy to describe. By normalising every long root of $\Frg$ to have squared length 2, we can always choose a basis $\{v_1,\ldots,v_n\}$ of $\Gamma_r$ such that $\{q_1 v_1,\ldots,q_n v_n\}$ is a basis of $\Gamma_r^\vee$ for some integers $\{q_i\}$. Doing so induces an isomorphism
\begin{gather}
    A_\tau \cong E_{q_1 \tau}\times\ldots\times E_{q_n\tau},\quad E_{q_i\tau} = \bbC/(\bbZ+q_i \tau \bbZ).
\end{gather}
For simply laced root systems $\Gamma_r=\Gamma_r^\vee$ so any basis of the root lattice will give an isomorphism between $A_\tau$ and $E_\tau^n$. When $\Frg$ is non-simply laced we can take the basis $\{v_i\}$ to be spanned by a choice of simple roots. Doing so gives the relative fibres shown in \cref{tab:relative}.

\begin{table}[]
    \centering
    \begin{tabular}{cccc} \hline\hline
        $\Frg$ & $d_{n-1}$ & $d_n$ & $A_\tau$ \\ \hline
        $A_n$ & $1$ & $n+1$ & $E_\tau^n$ \\
        $B_n$ & $1$ & $2$ & $E_\tau^{n-1}\times E_{2\tau}$ \\
        $C_n$ & $1$ & $2$ & $E_{2\tau}^{n-1}\times E_\tau$ \\
        $D_n$\, ($n$ odd) &$1$ & $4$ & $E_\tau^n$\\
        $D_n$\, ($n$ even) &$2$ &$2$ & $E_\tau^n$ \\
        $E_n$ & $1$ & $9-n$ & $E_\tau^{n}$\\
        $F_4$ & $1$ & $1$ & $E_{2\tau}^2\times E_\tau^2$ \\
        $G_2$ & $1$ & $1$ & $E_{3\tau}\times E_\tau$ \\ \hline \hline
    \end{tabular}
    \caption{The relative fibres for $\Frg$-type sYM along with their two largest invariant factors.}
    \label{tab:relative}
\end{table}

As $\cN=4$ theories are isotrivial \cite{Cecotti:2021ouq}, the generic smooth fibre of the relative ACI system is always isomorphic to $A_\tau$ and the total space of the fibration is
\begin{gather}
    \cX\simeq (\bbC^n\times A_\tau)/\cW[\Frg],
\end{gather}
where $\cW[\Frg]$ is the Weyl group of $\Frg$. Note that it acts on $\bbC^n$ via the (complexified) reflection representation and on $A_\tau$ via polarised automorphisms. Therefore, in order to correctly model $\cN=4$ sYM, $A_\tau$ must be endowed with a Weyl invariant polarisation. In \cite{Argyres:2022kon}, the authors showed that there is a natural Weyl invariant polarisation $\cL$ on $A_\tau$ whose alternating form (in a basis of simple roots and coroots) is given by
\begin{gather}\label{eq:cartan}
    E = \left[\begin{array}{cc}
         \mathbf{0} & C_\Frg \\
         -C_\Frg^T &  \mathbf{0}
    \end{array}\right],
\end{gather}
where $C_\Frg$ is the Cartan matrix of $\Frg$.\footnote{Note that this polarisation is related to the one constructed in \cite{EL76} for the varieties $\Gamma_r\otimes E_\tau$. Concretely, when $\Frg$ is simply laced they coincide, while the remaining cases are related by an appropriate isogeny.} The kernel of the corresponding isogeny then takes the form
\begin{gather}
    K(\cL) = (\Gamma_w\oplus\Gamma_w^\vee)/(\Gamma_r\oplus\Gamma_r^\vee),
\end{gather}
where $\Gamma_w$ and $\Gamma_w^\vee$ are the weight and coweight lattices respectively. It is well known that $\Gamma_w/\Gamma_r$ is isomorphic to the centre of the simply connected group $G$ associated to $\Frg$, while $\Gamma_w^\vee/\Gamma_r^\vee$ is dual to $\Gamma_w/\Gamma_r$ over $\bbQ/\bbZ$ and therefore isomorphic \cite{Bourb456}. As such, we have
\begin{gather}\label{eq:centre}
    K(\cL)\cong Z(G)\oplus Z(G)^*,
\end{gather}
where $Z(G)^*$ denote the Pontryagin dual of $Z(G)$. This is precisely the defect group expected of $\cN=4$ sYM \cite{Kapustin:2005py,DelZotto:2015isa,Gaiotto:2014kfa}. As such, the isogenies that produce the absolute fibres must have kernel of size $|Z(G)|$.

Finally, let us note that in general isogenies do not preserve the automorphism group of an abelian variety. However, as the resulting lattices we are interested in are sublattices of $\Gamma_w\oplus\Gamma_w^\vee$, it is easy to see that the Weyl group acts trivially on any subgroup of $K(\cL)$. Indeed, if $x\in\Gamma_w$ we have that any reflection $s_\alpha\in\cW[\Frg]$ acts as
\begin{gather}
    s_\alpha(x)=x-\langle x,\alpha^\vee\rangle\alpha.
\end{gather}
But by the definition of the weight lattice $\langle x,\alpha^\vee\rangle\in\bbZ$, so $s_\alpha(x) = x \ (\mathrm{mod}\ \Gamma_r)$. As such, the construction of the absolute fibres is entirely invariant under Weyl transformations and the automorphism group will still contain $\cW[\Frg]$.

\medskip
\noindent {\bf Total space structure.} As mentioned above, $\cN=4$ theories are isotrivial, meaning that every smooth fibre of the ACI system is isomorphic to a specified model fibre. Isotriviality descends to the absolute ACI systems as well; every smooth fibre is the same and we can apply the same $\cW[\Frg]$-equivariant isogeny to each relative fibre to get the absolute system. Furthermore, since the isogenies are $\cW[\Frg]$-equivariant, the total space description as a quotient still holds. That is, the total space $\cX'$ of the absolute system is given by
\begin{gather}
    \cX' \simeq (\mathbb{C}^r\times A'_\tau)/\cW[\Frg],
\end{gather}
where $A'_\tau$ is the absolute fibre and the $\cW[\Frg]$ action on $A'_\tau$ is specified by the original action on the relative fibre and the isogeny relating the two. Crucially, we see that whenever a point $c$ in $\mathbb{C}^r$ has a non-trivial stabiliser under the $\cW[\Frg]$ action, the fibre over $q(c)$, where $q$ is the quotient map, develops quotient singularities. As $\cW[\Frg]$ acts on the covering $\mathbb{C}^r$ via the reflection representation, Steinberg's theorem tells us that $c$ must belong to a reflection hyperplane \cite{MR117285,MR230728}. As such, the ACI system consists of smooth fibres away from the images of reflection hyperplanes, where singular fibres are obtained by taking the quotient of $A_\tau'$ by the reflection subgroups fixing those planes.

A further consequence of isotriviality is the fact that the holomorphic symplectic form $\Omega$ descends  from a $\cW[\Frg]$-invariant form $\hat{\Omega}$ in the cover $\mathbb{C}^r \times A_\tau$. In particular, we may always take it to be of the form\footnote{For more details about this, see the upcoming work \cite{isotriv} where the structure of general isotrivial theories are detailed more thoroughly.}
\begin{gather}
    \hat{\Omega} = \sum_{i=1}^r \diff x_i \wedge \diff z_i,
\end{gather}
where $\{x_i:i=1,\ldots,r\}$ are coordinates on $\mathbb{C}^r$ and $\{\diff z_i:i=1,\ldots,r\}$ are a basis of 1-forms on $A_\tau$. Here the $\diff z_i$ transform contragradiently to $\diff x_i$, meaning that $\hat{\Omega}$ is invariant under $\cW[\Frg]$ and therefore descends to a well-defined form on the quotient. Importantly, the $\diff z_i$ also transform under the analytic representation of an isogeny $f$ of the fibre $A_\tau$. As such, one can obtain the holomorphic 2-form of an absolute theory by first starting with $\hat{\Omega}$ for the relative theory, acting on $\diff z_i$ with the analytic representation of $f$ and taking the quotient by $\cW[\Frg]$.

These two points illustrate that the smooth fibre of an $\cN=4$ theory together and the quotient group $\cW[\Frg]$ completely determine the ACI system, regardless of whether it models a relative or absolute theory. For this reason, we now restrict attention to the smooth fibre herein.

\subsubsection{The pure refinements}\label{sec:universal}

As we know that each $Z(G)$ factor in \cref{eq:centre} is isotropic with respect to the natural polarisation on $A_\tau$, we can define two isogenies $f_1$ and $f_2$ such that
\begin{gather}
    \ker f_1 = Z(G)\oplus\mathbf{0},\quad \ker f_2= \mathbf{0}\oplus Z(G)^*
\end{gather}
At the level of the homology lattices, this is achieved by refining $\BPSL$ to $\Gamma_w+\tau \Gamma_r^\vee$ and $\Gamma_r+\tau\Gamma_w^\vee$ respectively. As remarked in \cite{Argyres:2022kon}, these can be identified with the $G$ and $(G/Z(G))_0$ global forms of the theory. As these are the `purely' electric and magnetic phases of the theory, we call these the \emph{pure refinements}.

Note that the integral dual of $\Gamma_r$ is precisely $\Gamma_w^\vee$, so if $\rho$ is the action of the Weyl group on $\Gamma_r$ then the dual (contragradiant) representation $\rho^*=\rho^{-T}$ acts on $\Gamma_w^\vee$. This gives the resulting fibre $\bbC^n/(\Gamma_r+\tau\Gamma_w^\vee)$ the structure of a $\rho$-decomposable variety in the language of \cite[def. 2.2]{CGR06}. 

\begin{defn}
    Let $X=\bbC^g/\Lambda$ be an abelian variety with principal polarisation $E=\mathrm{Im}\, H$ and $\mathcal{G}$ a finite group with a faithful integral representation $\rho$. If $\Lambda$ admits a splitting $\Lambda=L_1\oplus L_2$ into rank $g$ lattices $L_1$ and $L_2$ that are isotropic with respect to $E$ and are invariant under $\rho$ and the dual representation $\rho^*=\rho^{-T}$ respectively, then we say $X$ is $\rho$-decomposable.
\end{defn}

\begin{table}[t!]
    \centering
    \begin{tabular}{cc} \hline\hline
        $\Frg$ &  $\cB_\tau[\Frg]$ \\ \hline
        $A_n$ &  $E_\tau^{n-1}\times E_{\tau/(n+1)}$ \\
        $B_n$ &  $E_\tau^{n}$ \\
        $D_n$\, ($n$ odd) & $E_\tau^{n-1}\times E_{\tau/4}$ \\
        $D_n$\, ($n$ even) & $E_\tau^{n-2}\times E_{\tau/2}^2$ \\
        $E_n$ & $E_\tau^{n-1}\times E_{\tau/(9-n)}$ \\
        $F_4$ &  $E_{2\tau}^2\times E_\tau^2$ \\
        $G_2$ &  $E_{3\tau}\times E_\tau$ 
        \\ \hline \hline
    \end{tabular}
    \caption{The fibres $\cB_\tau[\Frg]=\bbC^n/(\Gamma_r+\tau\Gamma_w^\vee)$ for the magnetic phase of $\Frg$-type sYM. Here we omit the $C_n$ case, as the fibres coincide with the $D_n$ case after replacing $\tau$ with $2\tau$ (in our conventions).}
    \label{tab:magnetic}
\end{table}

\noindent Similarly, since $\Gamma_w^*= \Gamma_r^\vee$, the electric fibre $\bbC^n/(\Gamma_w+\tau\Gamma_r^\vee)$ is $\rho'$-decomposable where $\rho'$ is an integral representation extending the action $\rho$ on $\Gamma_r$ to $\Gamma_w$. The structure of $\rho$-decomposable varieties is incredibly constrained--- they all take the form $\bbC^g/(L+\tau L^*)$, for some $G$-invariant lattice $L$, and are isomorphic as complex tori to a product of elliptic curves \cite{CGR06}. Note that the pure refinements may not give the only $\rho$-decomposable varieties for a given $\Frg$. For example, when $\Frg=A_3$ there is an additional $\rho$-decomposable variety.

\Cref{tab:magnetic} lists the structure of the varieties $\cB_\tau[\Frg]=\bbC^n/(\Gamma_r+\tau \Gamma_w^\vee)$ as complex tori. The polarisation can then be deduced by considering the rational representations of the maps in \cref{eq:comm}. However, there is some overlap between the cases due to the fact that the varieties $\cB_\tau[\Frg]$ are invariant under all automorphisms of the corresponding root system and not just the Weyl group. In particular, for $\Frg=A_2$ or $D_n$, the automorphism group can be identified with the Weyl group of another root system, leading to the following identifications \cite{Feit98}:
\begin{gather}
    \begin{aligned}
        \Aut{A_2}&= \cW[G_2], &\quad \cB_{3\tau}[A_2]&\cong \cB_{\tau}[G_2],\\
        \Aut{D_4} &=\cW[F_4], &\quad \cB_{2\tau}[D_4]&\cong \cB_{\tau}[F_4],\\
        \Aut{D_n}&= \cW[C_n],  &\quad \cB_{2\tau}[D_n]&\cong \cB_{\tau}[C_n],\quad (n>4).
    \end{aligned}
\end{gather}
For the case $\Frg=G_2$, we will show this explicitly in \cref{ex:a2}. 

Given that the fibres for $B_8$ and $E_8$ are both given by $E_\tau^8$, one might also wonder if they coincide as abelian varieties. This can be easily seen to not be the case by noting that the complex torus $E_\tau^8$ admits only two distinct principal polarisations \cite{lange06}--- the product principal polarisation and an irreducible polarisation. Equipped with the product polarisation, the variety $E_\tau^8$ has automorphism group precisely $B_8$ when $\tau$ is generic. Since $B_8\nsubseteq E_8$, we conclude that $\cB_\tau[E_8]$ must be endowed with the irreducible polarisation and $\cB_\tau[B_8]$ must use the product polarisation. Therefore, the two fibres coincide as complex tori only.

\begin{remark}
    Despite the varieties in \cref{tab:magnetic} being reducible as complex tori, it is often the case that they are irreducible as polarised abelian varieties \cite{CGR06}.  We will discuss the consequences of this in \cref{sec:swcurves}.
\end{remark}

\subsubsection{Absolute fibres for \texorpdfstring{$A_\ell$}{A-type} sYM}

Let us now illustrate the how to obtain the absolute fibres via isogenies for $A$-type sYM. Here we follow the conventions of \cite{Bourb456} for the root and weight lattices.

First of all, note that the the root and weight lattice can be written as subsets of $\bbR^{\ell+1}$ as
\begin{gather}
    \Gamma_r=\bigoplus_{i=1}^\ell \bbZ(\epsilon_{i+1}-\epsilon_i), \quad \Gamma_w=\Gamma_r + \bbZ\omega,
\end{gather}
where $\{\epsilon_i\}$ is an orthonormal basis of $\bbR^{\ell+1}$ and $\omega= \epsilon_1-\tfrac{1}{(\ell+1)}(\epsilon_1+\epsilon_2+\ldots+\epsilon_{\ell+1})$. Furthermore, as all roots have squared length $2$ the root and coroot lattice coincide. As such, we have $\Gamma_w/\Gamma_r = \Gamma_w^\vee/\Gamma_r^\vee$ is generated by $\omega$ modulo the root lattice. The defect group in \cref{eq:centre} is therefore
\begin{gather}\label{eq:al_kernel}
    K(\cL) = \langle [\omega],\tau [\omega]\rangle \cong \bbZ_{\ell+1}\oplus \bbZ_{\ell+1}.
\end{gather}
In order to calculate the fibres for the absolute versions of this theory we must therefore quotient by subgroups of $K(\cL)$ of size $(\ell+1)$. It is known, see theorem 4 of \cite{znznsubgroups} for example, that the number of such subgroups is given by
\begin{gather}
    \sigma_1(N) = \sum_{k|N} k,
\end{gather}
matching the number of possible global forms/line lattices in \cite{Aharony:2013hda,Argyres:2022kon}.

As $\Gamma_r=\Gamma_r^\vee$, we have the freedom to choose any basis of $\Gamma_r$ to induce an isomorphism between the relative fibre $A_\tau$ and $E_\tau^\ell$. Noting that we can write $\omega$ in terms of the simple roots $\alpha_i$ as 
\begin{gather}
    (\ell+1)\omega = \sum_{i=1}^{\ell} (\ell-i+1)\alpha_i,\quad \alpha_i\doteq\epsilon_{i+1}-\epsilon_i,
\end{gather}
we see that we can take $\{\alpha_1,\ldots,\alpha_{\ell-1},(\ell+1)\omega\}$ as a basis of $\Gamma_r$. In this basis the polarisation takes the form
\begin{gather}
    E=\left[\begin{array}{cc}
        \ph{-}\mathbf{0} & C' \\
        -C' & \mathbf{0}
    \end{array}\right],\quad C'=\left[\begin{array}{cccccc|c}
        \ph{-}2 & -1 & \ph{-}0 & \cdots & \ph{-}0 &\ph{-}0 & \ell+1 \\
        -1 &\ph{-}2 & -1 & \cdots & \ph{-}0 &\ph{-}0 & 0\\
        \ph{-}0 & -1 & \ph{-}2 & \cdots &\ph{-}0 & \ph{-}0 & 0 \\
        \ph{-}\vdots & \ph{-}\vdots &\ph{-}\vdots  & \ddots & \ph{-}\vdots & \ph{-}\vdots & \vdots\\
        \ph{-}0 & \ph{-}0 & \ph{-}0 & \cdots & \ph{-}2 & -1 & 0\\
        \ph{-}0 & \ph{-}0 & \ph{-}0 & \cdots & -1& \ph{-}2 & 0\\ \midrule
        \ \ell+1& \ph{-}0 & \ph{-}0 & \cdots & \ph{-}0&\ph{-}0 & \ell(\ell+1)
    \end{array}\right].
\end{gather}
The benefit of decomposing $A_\tau$ in this way is that $K(\cL)$ is then identified with $(\ell+1)$-torsion points of the elliptic curve $E_\tau$ associated to the basis vector $(\ell+1)\omega$. Furthermore, one can explicitly check that all subgroups of size $(\ell+1)$ are isotropic using the polarisation above. As such, the absolute fibres for $A_\ell$ sYM always take the form $E_\tau^{\ell-1}\times (E_\tau/H)$ for some subgroup $H<K(\cL)$ of size $(\ell+1)$.\footnote{These groups $H$ coincide with the rank 2 lattices used in \cite{Amariti:2015dxa} from the geometric engineering perspective on global structures.}

Let us highlight the cases considered in \cref{sec:universal}. Consider the two subgroups of $K(\cL)$ given by $H_1=\langle [\omega]\rangle$ and $H_2=\langle\tau[\omega]\rangle$. At the level of the lattices these quotients implement the pure refinements; $H_1$ refines $\BPSL$ to $\Gamma_w+\tau\Gamma_r$ while the quotient by $H_2$ refines $\BPSL$ to $\Gamma_r+\tau\Gamma_w$. More specifically, the quotient by $H_1$ is given by
\begin{gather}
    A_\tau/H_1 = \bbC^\ell/(\Gamma_w+\tau\Gamma_r)\cong E_\tau^{\ell-1}\times E_{(\ell+1)\tau},
\end{gather}
while the quotient by $H_2$ gives the absolute fibre 
\begin{gather}
    A_\tau/H_2 = \bbC^\ell/(\Gamma_r+\tau\Gamma_w)\cong E_\tau^{\ell-1}\times E_{\tau/(\ell+1)},
\end{gather}
These can be identified with the fibres of the $\SU_{\ell+1}$ and $(\SU_{\ell+1}/\bbZ_{\ell+1})_0$ theories respectively. We will discuss how they are related in \cref{sec:sdual}. 

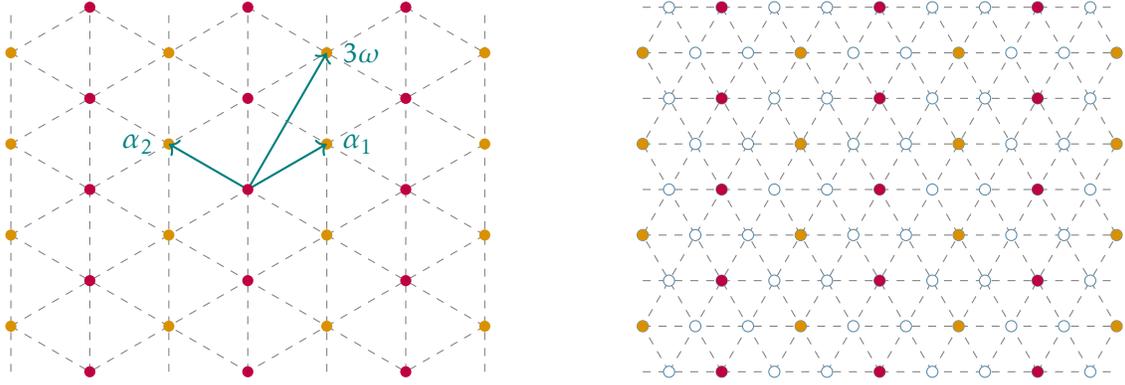
\begin{figure}
    \centering
    \begin{tikzpicture}[scale=0.7]
        
        \foreach \i in {-3,-2,...,3}{
            \draw[gray, dashed] (-1.5*\i, -4*0.866) -- (-1.5*\i, 4*0.866); 
        }
        \draw[gray,dashed] (-4.5,3*0.866) -- (4.5, -3*0.866);
        \draw[gray,dashed] (-4.5,3*0.866) -- (-3, 4*0.866);
        \draw[gray,dashed] (-4.5, 0.866) -- ( 3, -4*0.866);
        \draw[gray,dashed] (4.5,-0.866) -- (-3, 4*0.866);
        \draw[gray,dashed] (-4.5, 0.866) -- ( 0, 4*0.866);
        \draw[gray,dashed] (4.5, 0.866) -- ( -3, -4*0.866);
        \draw[gray,dashed] (-4.5, -0.866) -- ( 0, -4*0.866);
        \draw[gray,dashed] (-4.5, -0.866) -- ( 3, 4*0.866);
        \draw[gray,dashed] (4.5, -0.866) -- ( 0, -4*0.866);
        \draw[gray,dashed] (-4.5, -3*0.866) -- ( -3, -4*0.866);
        \draw[gray,dashed] (-4.5, -3*0.866) -- ( 4.5, 3*0.866);
        \draw[gray,dashed] (4.5, -3*0.866) -- ( 3, -4*0.866);
        \draw[gray,dashed] (0,4*0.866) -- (4.5, 0.866);
        \draw[gray, dashed] (3, 4*0.866) -- (4.5, 3*0.866);
                \foreach \i in {-1,0,...,2}{
            \foreach \j in {-2,-1,0,1}{
                \fill[harvestgold] (3*\i-1.5, 2*\j*0.866+0.866) circle (3pt);
            }
        }
        \draw[teal,thick,->] (0,0) -- (3/2,0.866);
        \node[teal,right] at (3/2+0.1,0.866){$\alpha_1$};
        \draw[teal,thick,->] (0,0) -- (-3/2,0.866);
        \node[teal,left] at (-3/2-0.1,0.866){$\alpha_2$};
        \draw[teal,thick,->] (0,0) -- (3/2,3*0.866);
        \node[teal,right] at (3/2+0.1, 3*0.866){$3\omega$};
        \foreach \i in {-2,-1,...,2}{
            \foreach \j in {-1,0,1}{
                \fill[purple] (3*\j,2*\i*0.866) circle (3pt);
            }
        }

        \begin{scope}[shift={(12,0)}]{
            \foreach \i in {-4,-3,...,4}{
                \draw[gray, dashed] (-4.5,\i*0.866) -- (4.5,\i*0.866);
            }
            \foreach \i in {-4,-3,...,0}{
                \draw[gray, dashed] (\i, -4*0.866) -- (\i+4, 4*0.866);
                \draw[gray, dashed] (-\i, -4*0.866) -- (-\i-4, 4*0.866);
            }
            \draw[gray, dashed] (1,  -4*0.866) -- (4.5,  3*0.866);
            \draw[gray, dashed] (2,  -4*0.866) -- (4.5,  1*0.866);
            \draw[gray, dashed] (3,  -4*0.866) -- (4.5,  -1*0.866);
            \draw[gray, dashed] (4,  -4*0.866) -- (4.5,  -3*0.866);
            \draw[gray, dashed] (1,  4*0.866) -- (4.5,  -3*0.866);
            \draw[gray, dashed] (2,  4*0.866) -- (4.5,  -1*0.866);
            \draw[gray, dashed] (3,  4*0.866) -- (4.5,  1*0.866);
            \draw[gray, dashed] (4,  4*0.866) -- (4.5,  3*0.866);
            \draw[gray, dashed] (-1,  -4*0.866) -- (-4.5,  3*0.866);
            \draw[gray, dashed] (-2,  -4*0.866) -- (-4.5,  1*0.866);
            \draw[gray, dashed] (-3,  -4*0.866) -- (-4.5,  -1*0.866);
            \draw[gray, dashed] (-4,  -4*0.866) -- (-4.5,  -3*0.866);       
            \draw[gray, dashed] (-1,  4*0.866) -- (-4.5,  -3*0.866);
            \draw[gray, dashed] (-2,  4*0.866) -- (-4.5,  -1*0.866);
            \draw[gray, dashed] (-3,  4*0.866) -- (-4.5,  1*0.866);
            \draw[gray, dashed] (-4,  4*0.866) -- (-4.5,  3*0.866);
            
            \foreach \i in {-4,-3,...,4}{
                \foreach \j in {-2,-1,...,2}{
                    \draw[airforceblue, fill=white] (\i,2*\j*0.866) circle (3pt);
                }
            }
            \foreach \i in {-5,-4,...,4}{
                \foreach \j in {-1,0,...,2}{
                    \draw[airforceblue, fill=white] (0.5+\i,2*\j*0.866-0.866) circle (3pt);
                }
            }
            \foreach \i in {-2,-1,...,2}{
                \foreach \j in {-1,0,1}{
                    \fill[purple] (3*\j,2*\i*0.866) circle (3pt);
                }
            }
            \foreach \i in {-1,0,...,2}{
                \foreach \j in {-2,-1,0,1}{
                    \fill[harvestgold] (3*\i-1.5, 2*\j*0.866+0.866) circle (3pt);
                }
            }
        }
        \end{scope}
    \end{tikzpicture}
    \caption{\emph{Left:} The root lattice $\Gamma_r\cong\langle\alpha_1,\alpha_2\rangle_\bbZ$ of $A_2$. \emph{Right:} A refinement of $\Gamma_r$ given by the weight lattice $\Gamma_w \cong \langle \alpha_1,\omega\rangle_\bbZ$. Highlighted are the nodes of $\Gamma_r$ present in $\Gamma_w$.}
    \label{fig:a2}
\end{figure}

\begin{example}\label{ex:su2}
    To start we shall consider the simplest example where $\Frg=A_1$. As this is a one dimensional Lie algebra, the generic fibre is simply an elliptic curve $A_\tau= E_\tau$, while the appropriate polarisation is 
    \begin{gather}
        E = \left[ 
            \begin{array}{cc}
                \ph{-}0 & \ph{-}2 \\
                -2 &  \ph{-}0
            \end{array}
        \right].
    \end{gather}
    It follows that the defect group is the set of 2-torsion points $E_\tau[2]=\langle 1/2,\tau/2\rangle$ and that the three maximally isotropic subgroups are 
    \begin{gather}
        H_1 = \langle 1/2\rangle,\quad H_2=\langle \tau/2\rangle,\quad H_3=\langle (1+\tau)/2\rangle.
    \end{gather}
   This results in the three absolute fibres
   \begin{gather}
       A_\tau/H_1 \cong E_{2\tau},\quad A_\tau/H_2\cong E_{\tau/2}, \quad A_\tau/H_3 \cong E_{(1+\tau)/2}.
   \end{gather}
   As an elliptic curve only admits one principal polarisation, all three absolute curves share this polarisation. Also note that the relation between the complex moduli of these curves and $\tau$ reproduces the expected relations between the relative and absolute couplings, as in \cite{Closset:2023pmc}.
\end{example}

\begin{example}\label{ex:a2}
    Let us now consider the $\Frg=A_2$ case. Let $\{\alpha_1,\alpha_2\}$ be a set of simple roots given in \cref{fig:a2} and $3\omega=2\alpha_1+\alpha_2$. In the basis $\{\alpha_1,-3\omega\}$ the polarisation on the relative fibre $A_\tau$ is given by
    \begin{gather}\label{eq:a2pol}
        E=\left[\begin{array}{cc}
        \ph{-}\mathbf{0} & C' \\
        -C' & \mathbf{0}
    \end{array}\right],\quad C'=\left[\begin{array}{cc}
        \ph{-}2 & -3 \\
        -3 & \ph{-}6
    \end{array}\right],
    \end{gather}
    while the defect group is given by $K(\cL)=\langle[\omega],\tau[\omega]\rangle\cong \bbZ_3\oplus \bbZ_3$. There are four maximally isotropic subgroups of $K(\cL)$ given by
    \begin{gather}
        H_1=\langle[\omega]\rangle,\ H_2=\langle\tau[\omega]\rangle,\ H_3=\langle[(\tau+1)\omega]\rangle,\ H_4=\langle[(\tau+2)\omega]\rangle.
    \end{gather}
    The quotients by $H_1$ and $H_2$ give the fibres discussed above, while the remaining quotients result in the fibres
    \begin{gather}
        A_\tau/H_3= E_{\tau}\times E_{(\tau+1)/3}, \quad A_\tau/H_4= E_{\tau}\times E_{(\tau+2)/3}.
    \end{gather}
    The polarisations on these fibres can then be easily inferred from \cref{eq:comm}. In particular, notice that $C'$ is the Killing form for a basis of simple roots for $G_2$. By performing the quotient by $H_1$ the resulting polarisation is then
    \begin{gather}
        E'=P^T\cdot E\cdot P=\left[\begin{array}{cccc}
             \ph{-}0& \ph{-}0 &\ph{-}2 & -3  \\
             \ph{-}0& \ph{-}0 & -1 & \ph{-}2 \\
             -2 & \ph{-}1 &\ph{-}0 &\ph{-}0 \\
             \ph{-}3 & -2&\ph{-}0 &\ph{-}0
        \end{array}\right],\quad P=\mathrm{diag}(1,1/3,1,1).
    \end{gather}
    which matches the polarisation in \cref{eq:cartan} for $\Frg=G_2$. Furthermore, the resulting complex torus $E_{3\tau}\times E_\tau$ matches the structure of the fibre expected for $G_2$, so we conclude that fibres for $\SU(3)$ sYM and $G_2$ sYM coincide as polarised abelian varieties.
\end{example}

\begin{remark}
    In this example all subgroups of $K(\cL)$ of the appropriate size were isotropic. We stress that this need not be the case when $K(\cL)\neq \bbZ_N\oplus\bbZ_N$. In particular, for $\Frg=D_n$ with $n$ even, the defect group is $(\bbZ_2\oplus\bbZ_2)\oplus(\bbZ_2\oplus\bbZ_2)$ giving $35$ possible subgroups of size $4$. However, it is easy to check that only 15 of these are isotropic with respect to $\e^\cL$.
\end{remark}

\subsubsection{Non-simply laced example}

In the $A$-type example, we saw the absolute fibres took the form of products of mutually isogenous elliptic curves due to the fact that the kernel of the polarisation took that form $\langle [\omega],\tau[\omega]\rangle$ with $(\ell+1)\omega\in\Gamma_r$. This is not always the case for other choices of gauge algebra. In particular, for $\Frg=B_n$ this fails, as we will show for $n=2$.

\begin{example}\label{ex:b2}
    For the root system $B_2\cong C_2$, we have that the relative fibre is given by
    \begin{gather}
        A_\tau = \bbC^2/\BPSL\cong E_{2\tau}\times E_\tau,
    \end{gather}
    with defect group given by $K(\cL)\cong \bbZ_2\oplus\bbZ_2$. However, while $\Gamma_w/\Gamma_r$ is isomorphic to $\Gamma_w^\vee/\Gamma_r^\vee$, the kernel does not take the form $\langle [\omega],\tau [\omega]\rangle$ as in the $A_\ell$ case. Indeed, the appropriate lattices are 
    \begin{gather}
        \begin{aligned}
            \Gamma_r&=\bbZ \epsilon_1\oplus \bbZ(\epsilon_2-\epsilon_1), &  \Gamma_r^\vee &= \bbZ(\epsilon_2-\epsilon_1)\oplus 2\bbZ\epsilon_1,\\
             \Gamma_w &=\bbZ \epsilon_1\oplus \tfrac{1}{2}\bbZ(\epsilon_2-\epsilon_1), & \Gamma_w^\vee &= \bbZ \epsilon_1\oplus \bbZ(\epsilon_2-\epsilon_1).
        \end{aligned}
    \end{gather}
    From this we see that the defect group takes the form $K(\cL) = \langle \tfrac{1}{2}(\epsilon_2-\epsilon_1),\tau\epsilon_1\rangle$. There are three maximally isotropic subgroups given by
    \begin{gather}
        H_1 = \langle \tfrac{1}{2}(\epsilon_2-\epsilon_1)\rangle,\quad H_2=\langle\tau\epsilon_1\rangle,\quad H_3=\langle\tfrac{1}{2}(\epsilon_2-\epsilon_1)+\tau\epsilon_1\rangle.
    \end{gather}
    The quotients by the first two give the familiar refinements to $\Gamma_w+\tau\Gamma_r^\vee$ and $\Gamma_r+\tau\Gamma_w^\vee$, with fibres given by
    \begin{gather}
        A_\tau/H_1\cong E_{2\tau}\times E_{2\tau},\qquad A_\tau/H_2 \cong E_\tau\times E_\tau.
    \end{gather}
    Note that for generic $z$, varieties of form $E_{z}\times E_z$ cannot be the Jacobian of a smooth genus 2 Riemann surface \cite{Haya65}. As such, the resulting polarisation is necessarily the product of the principal polarisations on each elliptic factor.

    In order to understand the quotient by $H_3$, it is useful to investigate how the period matrix of $A_\tau$ changes under the isogeny. We note that a symplectic basis for $\BPSL$ is given by
    \begin{gather}
        \lambda_1=\epsilon_1,\quad \lambda_2=\epsilon_2-\epsilon_1,\quad\lambda_3=2\tau\lambda_1+\tau\lambda_2,\quad \lambda_4=2\tau\lambda_1+2\tau\lambda_2. 
    \end{gather}
    By taking the basis of $V$ to be $\{\lambda_1,\tfrac{1}{2}\lambda_2\}$ we get the period matrix
    \begin{gather}\label{eq:pmat}
        \Pi=\left[
            \begin{array}{cc|cc}
                1 & 0 & 2\tau & 2\tau \\
                0 & 2 & 2\tau & 4\tau
            \end{array}
        \right].
    \end{gather}
    In this basis, $H_3$ can be written as $\langle\tfrac{1}{2}\lambda_4-\tfrac{1}{2}\lambda_2\rangle$. As such, a period matrix after isogeny is given by
    \begin{gather}
        \Pi' = G \Pi R = \left[
        \begin{array}{cc|cc}
            1 & 0 & 2\tau & \tau\\
            0 & 1 & \tau & \tau-\tfrac{1}{2}
        \end{array}
        \right],
    \end{gather}
    where
    \begin{gather}
        G=\left[\begin{array}{cc}
            1 & 0 \\
            0 & \tfrac{1}{2}
        \end{array}\right],\quad R=\left[
        \begin{array}{cccc}
            1 & 0 & 0 & \ph{-}0 \\
            0 & 1 & 0 & -\tfrac{1}{2} \\
            0 & 0 & 1 & \ph{-}0 \\
            0 & 0 & 0 & \ph{-}\tfrac{1}{2}
        \end{array}
        \right].
    \end{gather}
    We can bring $\Pi'$ into a more familiar form by further changing bases, as long as we preserve the polarisation. In particular, as the resulting polarisation on $J_\tau=A_\tau/H_3$ is principal, we must ensure any lattice basis changes are $\Sp_4(\bbZ)$ valued. Using the following $\Sp_4(\bbZ)$-valued basis change 
    \begin{gather}
        \tilde{G}=\left[\begin{array}{cc}
            1 & -1 \\
            0 & \ph{-}1
        \end{array}\right],\quad \tilde{R}=\left[
            \begin{array}{cccc}
                1 & 1 & \ph{-}0 &0\\
                0 & 1 & \ph{-}0 &0\\
                0 & 0 & \ph{-}1 & 0\\
                0 & 0 & -1& 1
            \end{array}
        \right],
    \end{gather}
    the period matrix takes the form
    \begin{gather}\label{eq:b2per}
        \tilde{\Pi}= \tilde{G}\Pi'\tilde{R}=\left[ 
            \begin{array}{cc|cc}
                1 & 0 & \tau-\tfrac{1}{2} & \tfrac{1}{2}\\
                0 & 1 & \tfrac{1}{2} & \tau-\tfrac{1}{2}
            \end{array}
        \right].
    \end{gather}
    By comparing this with \cite{MR369362}, we see that this is precisely the period matrix for the Jacobian of the curve $y^2=x(x^4+c(\tau)x^2+1)$.\footnote{This curve is equivalent to the automorphism frame curve found in \cite{Argyres:2023eij} through a change of variables.} Here $c(\tau)$ is a function relating the shape of the curve to the modulus of the Jacobian. When $c=0$ the automorphism group enhances to $\GL_2(\mathbb{F}_3)$, which corresponds to $\tau=\I/\sqrt{2}$.  As this is the Jacobian of a smooth curve for generic $c$, we see that the absolute fibre irreducible as a polarised abelian variety, unlike the other two global forms.
\end{example}

\subsection{\texorpdfstring{$S$}{S}-duality action on fibres}\label{sec:sdual}

\subsubsection{Relative conformal manifolds}

The relative fibres we have found in \cref{sec:relative} occur in 1-dimensional families $\cF=\{A_\tau\}$ indexed by the exactly marginal coupling $\tau\in\bbH_1$. However, as abelian varieties fall into $\Sp_{2r}^D(\bbZ)$ isomorphism classes, it is possible that $X_{\tau_1}\cong X_{\tau_2}$ for $\tau_1\neq\tau_2$. As such, the conformal manifold of the theory is actually $\bbH_1/\cS$ where $\cS$ is given by
\begin{gather}\label{eq:sduality}
    S = \{ f:\mathbb{H}_1\rightarrow \mathbb{H}_1: \exists M_f\in \Sp_{2r}^D(\bbZ) \text{ such that } M_f\cdot A_\tau \cong A_{f(\tau)},\, \forall\tau\in\mathbb{H}_1\}.
\end{gather}
In other words, $\cS$ consists of transformations acting on $\tau$ that lead to isomorphisms amongst the whole family $\cF$. Physically, the group $\cS$ is known as the $S$-\emph{duality group} and gives rise to exact equivalences of theories with differing couplings. Let us show that the varieties we have constructed thus far reproduce the expected $S$-duality groups.

When $\Frg$ is simply laced, the $S$-duality group is easy to deduce. Recall that the relative fibres are given by $\Gamma_r\otimes E_\tau\cong E_\tau^\ell$ for some $\ell\in\bbN$ and an elliptic curve $E_\tau$ is isomorphic to another $E_{\tau'}$ if and only if
\begin{gather}
    \tau'=\frac{a\tau+b}{c\tau+d},\quad\quad \left[\begin{array}{cc} 
        a & b \\
        c & d
    \end{array}\right]\in\PSL_2(\bbZ).
\end{gather}
As such, this action is inherited by $A_\tau$ and we see the family $\cF=\{A_\tau\}$ is preserved by this $\PSL_2(\bbZ)$ action. Furthermore, as the polarisation is independent of $\tau$, we conclude that there is a corresponding $\Sp_{2r}^D(\bbZ)$-valued transformation representing this $\PSL_2(\bbZ)$ action, thus making it an isomorphism of abelian varieties. As the fibres in question are products of elliptic curves, it is clear that there are no other $\Sp_{2r}^D(\bbZ)$ transformations which preserve the form of the family $\cF$, so we conclude that the conformal manifold of these relative theories is the modular curve $X(1)=\bbH_1/\PSL_2(\bbZ)$. 

For $\Frg$ non-simply laced, the family $\cF$ is no longer invariant under the full modular group $\PSL_2(\bbZ)$. This can be seen by noting that the fibres in this case take the form $A_\tau = E_{q\tau}^m\times E_\tau^n$ for integers $m,n$ and $q$. Now the action of $S:\tau\mapsto -1/\tau$ maps the fibre to
\begin{gather}
    A_\tau\mapsto E_{-q/\tau}^m \times E_{-1/\tau}^n \cong E_{\tau/q}^m \times E_{\tau}^n,
\end{gather}
where we have used the $\PSL_2(\bbZ)$ invariance of elliptic curves in the last step. Therefore, $S\in \PSL_2(\bbZ)$ does not stabilise the family $\cF$. However, one can check that $S T^q S$, where $T:\tau\mapsto\tau+1$, \emph{does} preserve $\cF$. In particular, $S T^q S : \tau\mapsto \tau/(1-q\tau)$ so $A_\tau$ transforms as
\begin{gather}
    A_\tau \mapsto E_{q\tau/(1-q\tau)}^m \times E_{\tau/(1-q\tau)}^n.
\end{gather}
As $S T^q S\in \PSL_2(\bbZ)$, the second factor is clearly invariant under this action, while the invariance of the first factor can be established by noting
\begin{gather}
    \left[\begin{array}{cc}
        1 & 0 \\
        1 & 1
    \end{array}\right]: \frac{q\tau}{1-q\tau}\mapsto q\tau,
\end{gather}
under the usual fractional linear action. We therefore conclude that $A_\tau$ is invariant under the group generated by $T$ and $S T^q S$ --- the so-called \emph{modular congruence subgroup} $\Gamma_0(q)$ of $\PSL_2(\bbZ)$. These exhaust the $\PSL_2(\bbZ)$ transformations which stabilise $\cF$.

While these varieties are no longer invariant under all of $\PSL_2(\bbZ)$, there are additional transformations that arise from elements of $\PSL_2(\bbR)$ to consider \cite{Girardello:1995gf,Dorey:1996hx,Argyres:2006qr}. These additional isomorphisms can enhance the congruence subgroups above to \emph{Hecke triangle groups}. To obtain these groups from $\PSL_2(\bbZ)$, we replace the generator $S:\tau\mapsto -1/\tau$ with $S_\ell:\tau\mapsto -1/\ell\tau$ where $\ell\in\{2,3\}$ \cite{MR1513069}. Doing so gives the group presentation
\begin{gather}
    \cH_q = \langle S_\ell, T: S_\ell^2=(S_\ell T)^q=\mathrm{id}\rangle \cong \bbZ_2 *\bbZ_q,
\end{gather}
where $\ell=4\cos^2(\pi/q)$. Under such a transformation, the fibre $A_\tau$ transforms as
\begin{gather}
    S_q:A_\tau \mapsto E_{-1/\tau}^m \times E_{-1/q\tau}^n \cong E_{\tau}^m \times E_{q\tau}^n,
\end{gather}
from which we see that $A_\tau$ is invariant (as a complex torus) if and only if $m=n$. By comparing with \cref{tab:relative} we see that this is the case for $\Frg = F_4, G_2$ and $B_2$. For $B_n$ and $C_n$ with $n\geq 3$, we see that the two fibres are exchanged under $S_2$, which is expected as $B_n = \lang{C_n}$. 

Despite the Hecke triangle groups containing transformations corresponding to $\PSL_2(\bbR)$ actions on $\tau$, one can still construct an $\Sp_{2r}^D(\bbZ)$-valued action on $\bbH_r$. To illustrate this, let us return to the $B_2$ case. Under the transformation $S_2$, the period matrix in \cref{eq:pmat} becomes
\begin{gather}
    S_2 \cdot \Pi = \left[
            \begin{array}{cc|cc}
                1 & 0 & -\tfrac{1}{\tau} & -\tfrac{1}{\tau} \\
                0 & 2 & -\tfrac{1}{\tau} & -\tfrac{2}{\tau}
            \end{array}
        \right].
\end{gather}
However, we have that $G(S_2\cdot\Pi)R_S=\Pi$ with
\begin{gather}
    G=\left[\begin{array}{cc}
         -2\tau& \tau \\
         -2\tau & 0
    \end{array}\right],\quad R_S=\left[\begin{array}{cccc}
        \ph{-}0 & \ph{-}0 & -1&-2 \\
        \ph{-}0 & \ph{-}0 & \ph{-}0&-1\\
        \ph{-}1 & \ph{-}0&\ph{-}0&\ph{-}0\\
        -1&\ph{-}1&\ph{-}0&\ph{-}0
    \end{array}\right].
\end{gather}
where we notice that $R_S\in \Sp_4^{D}(\bbZ)$ with $D=\mathrm{diag}(1,2)$. Similarly, for $T:\tau\mapsto\tau+1$ we have the following relation
\begin{gather}
    (T^{-1}\cdot\Pi)R_T = \Pi,\quad R_T=\left[\begin{array}{cccc}
        1 & 0 & 2 & 2 \\
        0 & 1 & 1 & 2 \\
        0 & 0 & 1 & 0 \\
        0 & 0 & 0 & 1
    \end{array}\right],
\end{gather}
where, again, $R_T\in \Sp_4^{D}(\bbZ)$. We therefore conclude that $A_\tau$ is invariant under all of $\cH_4$. It is worth remarking that the group $\langle R_S, R_T\rangle\subset \Sp_4^{D}(\bbZ)$ is not isomorphic to $\cH_4$ as $R_S^2$ is non-trivial. However, $R_S^2$ represents an automorphism of $A_\tau$, so the effective action on $A_\tau$ is given by $\langle R_S,R_T\rangle/\langle R_S^2\rangle\cong\cH_4$. One can check that there are analogous $\Sp_{2r}^D(\bbZ)$ transformations representing the Hecke transformations found for the other non-simply laced cases. All in all, we conclude that the conformal manifolds are given by $X_0(2)=\bbH_1/\Gamma_0(2)$ for $\Frg=B_n, C_n$ with $n\geq 3$, $W(4)=\bbH_1/\cH_4$ for $\Frg= B_2, F_4$ and $W(6) =\bbH_1/\cH_6$ for $\Frg=G_2$.

\subsubsection{Absolute \texorpdfstring{$S$}{S}-duality groups}

When we pass from the relative theory to an absolute theory by specifying a spectrum of lines, it is known that the $S$-duality group can partially break \cite{Aharony:2013hda}. In our picture, this is reflected in the fact that $\cS$ gives rise to a non-trivial action on the maximally isotropic subgroups of $K(\cJ)$ which then descends to the absolute fibres after isogeny. We first exhibit this in the simple case of $\Frg=A_n$, before making some more general comments.\footnote{We do not aim to be systematic here, instead giving several examples to illustrate our fibres give the correct $S$-duality groups. For a more systematic approach, see \cite{SKLines}.}

\begin{example}\label{ex:su2sdual}
    For $\Frg=A_1$, we saw in \cref{ex:su2} that the relative fibre is the elliptic curve $E_\tau$ with two times the unique principal polarisation, leading to three absolute fibres. The relative $S$-duality group is therefore $\PSL_2(\mathbb{Z})$ with the usual fractional linear action on $\tau$. Using this action on the absolute fibres we see that $T:\tau\mapsto \tau+1$ leaves $E_{2\tau}$ invariant and interchanges $E_{\tau/2}$ and $E_{(1+\tau)/2}$, while $S:\tau\mapsto-1/\tau$ fixes $E_{(1+\tau)/2}$ and interchanges $E_{2\tau}$ and $E_{\tau/2}$. This leads to the following $S$-duality diagram.
    \begin{gather*}
        \xymatrix{
            E_{2\tau} \ar@{<->}[r]^{S}\ar@(ul,ur)^{T} & E_{\tau/2} \ar@{<->}[r]^{T} & E_{(1+\tau)/2} \ar@(ul,ur)^{S}
        }
    \end{gather*}
    Each form is no longer invariant under all of $\SL_2(\mathbb{Z})$, instead the relative $S$-duality group is broken to a subgroup conjugate to $\Gamma_0(2)$, matching the expectations of \cite{Aharony:2013hda}.
\end{example}

\begin{example}
    We now move to the $\SU_3$ case. Recall that there are 4 maximally isotropic subgroups of $K(\cJ)\cong \bbZ_3\oplus\bbZ_3$ which result in the four absolute fibres discussed in \cref{ex:a2}. Starting with $A_\tau/H_1 \cong E_\tau\times E_{3\tau}$, we see that it is invariant under $T:\tau\mapsto \tau+1$, but under the action of $S:\tau\mapsto -1/\tau$ we obtain
    \begin{gather}
        A_\tau/H_1 = E_\tau\times E_{3\tau}\mapsto E_{-1/\tau}\times E_{-3/\tau} \cong E_\tau\times E_{\tau/3}.
    \end{gather}
    By identifying this with $A_\tau/H_2$ and we see that $S$ exchanges the purely electric fibre with the purely magnetic one. This signals that the $S$-duality group has broken to a subgroup not including $S$--- specifically, one can check that these global forms are only $\Gamma_0(3)$ invariant.   
    
    We can continue by acting with $T$ and $T^2$ on $A_\tau/H_2$ to obtain the final two global forms. These two forms are also related by $S$, which can be seen by noting that $S$ maps $A_\tau/H_3$ to $E_{-1/\tau}\times E_{(\tau-1)/3\tau}$. This is isomorphic to $A_\tau/H_4$ since
    \begin{gather}
        \left[\begin{array}{cc}
            2 & -1 \\
            3 & -1
        \end{array}\right]: \frac{\tau-1}{3\tau}\mapsto \frac{\tau+2}{3}.
    \end{gather}
    Therefore, all global forms fall into a single $S$-duality orbit with residual $S$-duality group $\Gamma_0(3)$. This can be summarised as below.
    \begin{gather*}
        \xymatrix@R=1pc{
         & & E_\tau\times E_{(\tau+1)/3} \ar@/_/[dd]_{T} \ar@{<->}@/^/[dd]^{S}\\
        E_\tau\times E_{3\tau} \ar@(ul,ur)^{T} \ar@{<->}[r]^{S} & E_\tau\times E_{\tau/3} \ar[ur]^{T} \\
        & & E_\tau\times E_{(\tau+2)/3} \ar[ul]^{T}
        }
    \end{gather*}
\end{example}

\begin{example}
    In the $\SU_3$ case there was only $\SL_2(\bbZ)$ orbit, but this is not generic. As an example, consider the $\SU_4$ case. If we start with the purely electric fibre, as before, we can act with combinations of $S$ and $T$ to obtain an $S$-duality orbit containing 6 global forms with residual $S$-duality group $\Gamma_0(4)$. However, there is an additional global form which arises from taking the unique $\bbZ_2\oplus\bbZ_2$ subgroup of $K(\cJ)\cong \bbZ_4\oplus \bbZ_4$. This corresponds to the 2-torsion points $E_\tau[2]$ of one of the elliptic factors, so we obtain the absolute fibre
    \begin{gather}
        A_\tau/(\bbZ_2\oplus\bbZ_2) = E_\tau^2\times (E_\tau/E_\tau[2])\cong E_\tau^3.
    \end{gather}
    Furthermore, the polarisation is completely reducible, so the fibre is isomorphic to a product of elliptic curves as a polarised abelian variety. From this it is easy to see that this global form is invariant under all of $\PSL_2(\bbZ)$ and therefore forms a singlet under $S$-duality. The overall duality structure is summarised below.
    \begin{gather*}\scalebox{0.9}{
        \xymatrix@R=0.75pc{
         & & E_\tau^2\times E_{(\tau+1)/4} \ar[dr]_{T} \ar@{<->}[dd]^{S}\\
        E_\tau^2\times E_{4\tau} \ar@(ul,ur)^{T} \ar@{<->}[r]^{S} & E_\tau^2\times E_{\tau/4} \ar[ur]^{T}  & & E_\tau^2 \times E_{(\tau+2)/4}\ar[dl]_{T} \ar@{<->}[r]^{S} &  \ar@(ul,ur)[]^{T} E_\tau^2\times E_{(2\tau+1)/2} &  E_\tau^3 \ar@{<->}@(ul,ur)^{S} \ar@(dl,dr)_{T}\\
        & & E_\tau^2\times E_{(\tau+3)/4}  \ar[ul]^{T} 
        }}
    \end{gather*}
    Note that while the singlet has the product principal polarisation, the other $S$-duality orbit consists of irreducible abelian varieties.
\end{example}

The $A_3$ example highlights that the global forms of a theory do not have to be related via the $S$-duality group. In this case, the occurrence of distinct $S$-duality orbits can be traced to the fact that there are 7 maximally isotropic subgroups of $K(\cJ)$, six of which are isomorphic to $\bbZ_4$ and only one is isomorphic to $\bbZ_2\oplus\bbZ_2$. For $\Frg=A_n$ this generalises easily; the number of $S$-duality orbits is equal to the number of non-isomorphic maximally isotropic subgroups of $K(\cJ)\cong \bbZ_{n+1}\oplus\bbZ_{n+1}$. For example, if $n+1=p^{m}$ for a prime $p$, the maximally isotropic subgroups take the form $\bbZ_{p^{m-k}}\oplus \bbZ_{p^k}$. Accounting for the symmetry between the two cyclic factors, we see that there are $1+\lfloor m/2\rfloor$ many non-isomorphic subgroups. Extending this result to composite $p$, we have that if $n+1=\prod_i p_i^{m_i}$ is the prime decomposition of $n+1$, the number of distinct $S$-duality orbits is given by
\begin{gather}
    \Sigma(n+1) = \prod_i (1+\lfloor m_i/2\rfloor).
\end{gather}
This is precisely the number of square divisors of $n+1$, in agreement with the result of \cite{Bergman:2022otk}.

A benefit of using isogenies to produce the absolute fibres is that the complex moduli of the resulting fibre is written in terms of the exactly marginal coupling $\tau$ regardless of whether the resulting variety is completely split or not, thus allowing us to efficiently probe the residual $S$-duality of the fibre. To illustrate this, let us return to the $B_2$ case.

\begin{example}
    For $\Frg=B_2=C_2$ the two pure refinements give the fibres $E_\tau \times E_\tau$ and $E_{2\tau}\times E_{2\tau}$ which are exchanged under $\tau\mapsto -1/2\tau$. More interesting is the variety $J_\tau$, which we showed is the Jacobian of $y^2=x(x^4+c (\tau) x^2+1)$.
     
    Recall that the period matrix of $J_\tau$ is given by 
    \begin{gather}
        \Pi=\left[\begin{array}{cc|cc}
                1 & 0 & \tau-\tfrac{1}{2} & \tfrac{1}{2}\\
                0 & 1 & \tfrac{1}{2} & \tau-\tfrac{1}{2}
            \end{array}
        \right].
    \end{gather}
    First of all, notice that we can implement $T:\tau\mapsto \tau+1$ through the following operation on the period matrix
    \begin{gather}
        \left[\begin{array}{cc|cc}
                1 & 0 & \tau-\tfrac{1}{2} & \tfrac{1}{2}\\
                0 & 1 & \tfrac{1}{2} & \tau-\tfrac{1}{2}
            \end{array}
        \right]\left[\begin{array}{cccc}
            1 & 0 & 1 & 0 \\
            0 & 1 & 0 & 1 \\
            0 & 0 & 1 & 0 \\
            0 & 0 & 0 & 1
        \end{array}\right]=\left[\begin{array}{cc|cc}
                1 & 0 & \tau+\tfrac{1}{2} & \tfrac{1}{2}\\
                0 & 1 & \tfrac{1}{2} & \tau+\tfrac{1}{2}
            \end{array}
        \right]= T\cdot\Pi,
    \end{gather}
    establishing that the isomorphism class of $J_\tau$ is invariant under $T$.
    Similarly, under $S_2$, we have that $\tau\mapsto -1/(2\tau)$ and $\Pi\mapsto S_2\cdot \Pi$. However, this can also be undone by a change of bases as follows
    \begin{gather}
         G \left[\begin{array}{cc|cc}
                1 & 0 & -\tfrac{1}{2\tau}-\tfrac{1}{2} & \tfrac{1}{2}\\
                0 & 1 & \tfrac{1}{2} & -\tfrac{1}{2\tau}-\tfrac{1}{2}
            \end{array}\right] R= \left[\begin{array}{cc|cc}
                1 & 0 & \tau-\tfrac{1}{2} & \tfrac{1}{2}\\
                0 & 1 & \tfrac{1}{2} & \tau-\tfrac{1}{2}
            \end{array}
        \right],
    \end{gather}
    where the matrices $G$ and $R$ are given by
    \begin{gather}
    G=\left[\begin{array}{cc}
             \tau& -\tau  \\
             \tau& \ph{-}\tau 
        \end{array}\right],\quad
        R = \left[
            \begin{array}{cccc}
                -1 & \ph{-}0 & \ph{-}1 & 0\\
                 \ph{-}1& \ph{-}0 &-1 &1 \\
                 -1 & -1 & \ph{-}0 & 0 \\
                 \ph{-}1 & -1 & -1 & 1
            \end{array}
        \right].
    \end{gather}
    As $R$ is $\Sp_4(\bbZ)$-valued, this is a valid isomorphism of abelian varieties and we see that $J_{-1/2\tau}\cong J_\tau$. We therefore conclude that $J_\tau$ is self-dual under all of $\cH_2$. The full $S$-duality structure of the $B_2$ fibres can be summarised as below.
    
    \begin{gather*}
        \xymatrix{E_\tau^2 \ar@(ul,ur)^{T}\ar@{<->}[rr]^{S_2} & &E_{2\tau}^2\ar@(ul,ur)^{T} & & J_\tau \ar@(ul,ur)^{T}\ar@(dl,dr)_{S_2}}
    \end{gather*}
    While this matches the overall $S$-duality structure presented in \cite{Aharony:2013hda}, it does not match the prediction of \cite{Argyres:2023eij} that the non-split variety $J_\tau$ belongs to the doublet. Nevertheless, the isogeny point of view explicitly relates the complex moduli of the absolute variety to that of the relative variety, thus giving us direct confirmation of the above structure.
\end{example}

\begin{remark}
    Note that the definition of the $S$-duality group given in \cref{eq:sduality} does not necessarily hold for absolute theories. For example, if we consider the electric fibre for the $D_4$ case, we have $\cE_\tau= E_\tau^2 \times E_{2\tau}^2$ and we could consider the Hecke transformation $\tau\mapsto-1/2\tau$. This gives rise to an isomorphism of $\cE_\tau\cong \cE_{-1/2\tau}$, but we should \emph{not} think of this an element of the $S$-duality group. Indeed, this transformation is not an element of the relative $S$-duality group so we should disregard it. Physically, this would be a transformation that does not distinguish between the charges of genuine BPS states of the theory and charges occupied by screened lines only. If we wish to amend the definition in \cref{eq:sduality} for absolute theories, we must ensure that we only consider transformations of $\tau$ belonging to the relative duality group $\cS$. That is, we can define the absolute $S$-duality group to be
    \begin{gather}
        \cS_{\mathrm{abs}} = \{ f\in\cS: \exists M_f\in \Sp_{2r}(\bbZ) \text{ such that } M_f\cdot A_\tau' \cong A_{f(\tau)}',\, \forall\tau\in\mathbb{H}_1\}.
    \end{gather}
    where $A'_\tau$ is the absolute fibre in question.
\end{remark}

\section{Compatibility with Seiberg-Witten systems}\label{sec:swcurves}

While the ACI system associated to an $\cN\geq 2$ theory provides a uniform characterisation of the low energy dynamics of the theory, it is often more convenient to work with the more ubiquitous Seiberg-Witten system \cite{Seiberg:1994aj,Seiberg:1994rs}. Instead of working with abelian varieties, one is instead able to work with Riemann surfaces where simpler algebraic equations are attainable. However, it is important to note that the physical data at any point on the Coulomb branch does not necessarily define a Riemann surface, so this opens up the possibility that an absolute theory does not admit a Seiberg-Witten system. In this section, we comment upon the compatibility between the fibres in \cref{tab:magnetic} with possible Seiberg-Witten systems, focussing on the case of minimal genus solutions. We then compare this with the known higher genus solutions arising from the massless limit of (twisted) Calogero-Moser systems \cite{DHoker:1997hut,DHoker:1998xad}.

\subsection{Minimal genus compatibility}

For concreteness, we shall start by defining what we mean by a Seiberg-Witten system for our purposes. Given a rank $g$ theory, a Seiberg-Witten (SW) system for that theory is a fibration of Riemann surfaces $\cC_u$ of at least genus $g$ over the Coulomb branch together with a meromorphic $1$-form $\lambda_{\mathrm{SW}}\in\Omega^1(\cC_u)$ such that the special coordinates of the Coulomb branch and the effective couplings are given by 
\begin{gather}
    a_i(u)=\int_{\alpha_i} \lambda_{\mathrm{SW}},\quad 
    a^D_i(u)=\int_{\beta_i} \lambda_{\mathrm{SW}},\quad \tau_{ij}=\frac{\partial a^D_i}{\partial a_j},
\end{gather}
where $\{\alpha_i,\beta_i:i=1,\ldots, g\}$ are a set of 1-cycles spanning a rank $2g$ sublattice $\Lambda$ of $H_1(\cC_u,\bbZ)$ satisfying $\langle \alpha_i,\beta_j\rangle=d_i \epsilon_{ij}$ with $d_{i}|d_{i+1}$. The Dirac pairing of theory is then given by the intersection pairing on $\cC_u$ restricted to $\Lambda$, making the $1$-cycles $\{\alpha_i,\beta_i\}$ a symplectic basis of the Dirac pairing. As the special coordinates determine the metric on the Coulomb branch, the curves $\cC_u$ must develop vanishing cycles precisely at the singular points of the Coulomb branch. Encircling a component of the singular locus causes the elements of $H_1(\cC_u,\bbZ)$ to undergo a Picard-Lefschetz transformation that identifies which cycle vanishes at that point. Physically, this tells us the electromagnetic charges of the corresponding BPS state, so we further require that a valid Seiberg-Witten system reproduces the correct BPS spectrum.

Consider the case where the genus of $\cC_u$ is equal to the rank of the theory. In general, the homology lattice $H_1(\cC_u,\bbZ)\cong \bbZ^{2g}$ contains the BPS charge lattice, but does not necessarily equal it. Nevertheless, the intersection pairing provides a principal polarisation on $\jac(\cC_u)= \bbC^g/H_1(\cC_u,\bbZ)$ and we can naturally form an ACI system by replacing the fibre $\cC_u$ by its Jacobian and endowing it with the holomorphic two-form $\diff \lambda_{\mathrm{SW}}$. Since the polarisation on the fibres is principal, this signals that this system potentially models an absolute theory with $H_1(\cC_u,\bbZ)$ being identified with a refinement of the BPS charge lattice by probe lines. Therefore, given a valid SW system of minimal genus, this procedure produces a possible ACI system that describes the dynamics of the theory. However, if we are instead given an ACI system with principal polarisation, it is not guaranteed that there exists a fibration of Riemann surfaces whose Jacobians recover the integrable system in question due to the Schottky problem. This leads us to the following definition.

\begin{defn}
    Given an ACI system for an absolute theory with $g$-dimensional fibres $(A_u,\cL)$ and holomorphic 2-form $\Omega$, we say there is a compatible {\it minimal Seiberg-Witten system} if there exist genus $g$ Riemann surfaces $\cC_u$ and a meromorphic 1-form $\lambda$ such that 
    \begin{gather}
        (\jac(\cC_u),\Theta_{\cC_u}) \cong (A_u,\cL),\quad\quad \Omega = \diff \lambda,
    \end{gather}
    where $\Theta_{\cC_u}$ is the canonical polarisation on $\jac(\cC_u)$, for every smooth point $u$ in the Coulomb branch.
\end{defn}

\noindent With this definition in mind, two remarks are in order.
\begin{enumerate}
    \item Torelli's theorem tells us that a curve $\cC$ is determined by its Jacobian with its canonical polarisation $\Theta_\cC$. With the above the definition, if an absolute theory admits a minimal Seiberg-Witten system then it is completely determined by the ACI system and vice versa.
    \item Instead of forming the fibres of the ACI system by taking the Jacobian of $\cC_u$, one could ask about using just the sublattice of $H_1(\cC_u,\bbZ)$ occupied by BPS states to form the abelian fibres. If the theory admits non-trivial one-form symmetries, this would simply result in the relative ACI system which is isogenous to the fibration by Jacobians. Furthermore, since we could repeat this for each SW system corresponding to each global form, there would be no uniqueness as in the absolute case. It is for this reason we only consider the Jacobian fibrations, as the full homology lattice uniquely provides vital information about the global form of the theory.
\end{enumerate}


The Schottky problem captures the obstruction to a minimal SW system. In particular, if one knows that a fibre $A_u$ of the ACI system is an irreducible principally polarised abelian variety, then there exists a minimal Seiberg-Witten system if and only if $A_u$ belongs to the Torelli locus for all $u\in\cC$. More generally, one can allow for reducible fibres and SW curves which are bouquets of curves of lower genera, as in \cite{Argyres:2024hdn}, in which case an analysis of each irreducible factor is required and one must consider the Schottky problem for each. Regardless of reducibility, as we know that all principally polarised abelian varieties of dimension less than or equal to 3 are the Jacobians of a (possibly reducible) Riemann surface, we know that there will exist a minimal SW curve in theories of rank 3 and lower.\footnote{Here we are being a little cavalier. In general, it is not clear if all required features of a SW system can be reconstructed from a proposed ACI system. For example, while the corresponding local systems of both systems will encode the same monodromy representation, there may be no canonical way to obtain the singular SW curves from the singular fibres of the ACI system. We are therefore only considering the obstruction arising from the Schottky problem.} However, in rank 4 and above we no longer have this guarantee.

\subsection{Compatibility for \texorpdfstring{$\cN=4$}{N=4} phases}

\subsubsection{Irreducibility and automorphisms}

Let us turn our attention to the compatibility question for the varieties presented in \cref{tab:magnetic}. As these model the magnetic phase of an $\cN=4$ theory, every smooth fibre of the system is isomorphic to the variety given. So if one fibre is the Jacobian of a curve $C$, then Torelli's theorem states that the corresponding curve over any other smooth point is isomorphic to $C$. The question of compatibility is therefore reduced to understanding whether the varieties in \cref{tab:magnetic} lie in the Torelli locus $\cT_g$ or its closure.

\begin{table}[t]
    \centering
    \begin{tabular}{ccccc}\hline\hline
        $\Phi$ & $\cW[\Phi]$ & $|\cW[\Phi]|$ & Coxeter diagram  & $\dim P$ \\ \hline
        $A_n$ & $S_{n+1}$ & $(n+1)!$ & \dynkin[Coxeter]A{} & $n+1$\\
        $B_n$, $C_n$& $\bbZ_2\wr S_n$ & $2^n n!$ & \dynkin[Coxeter]B{} & $2n$ \\
        $D_n$ & $\bbZ_2^{n-1}\cdot S_n$ & $2^{n-1} n!$& \dynkin[Coxeter]D{} & $2n$\\
        $E_6$ & $\SO_5(\bbF_3)$& $2^7\cdot 3^4\cdot 5$& \dynkin[Coxeter]E{6} & $27$\\
        $E_7$ & $\bbZ_2\times \Sp_6(\bbF_2)$& $2^{10}\cdot 3^4\cdot 5\cdot 7$& \dynkin[Coxeter]E{7} & $56$\\
        $E_8$ & $\bbZ_2\cdot \mathrm{O}^+_8(\bbF_2)$ & $2^{14}\cdot 3^5\cdot 5^2\cdot 7$ & \dynkin[Coxeter]E{8} & $240$ \\ 
        $F_4$ & $\mathrm{O}^+_4(\bbF_3)$& $2^7 \cdot 3^2$ & \dynkin[Coxeter]F{4} &$24$ \\
        $G_2$ & $\dih{6}$ & $2^2\cdot 3$& \dynkin[Coxeter,gonality=6]G2 & $6$ \\
        \hline\hline
    \end{tabular}
    \caption{Information about the Weyl groups of semi-simple root systems. Here $P$ is the lowest dimensional faithful permutation representation and $\mathrm{O}^+$ is a form of the orthogonal group that occurs over fields of positive characteristic. See \cite[section 3.7]{wilson09} for more details. }
    \label{tab:coxeter}
\end{table}

As reviewed in \cref{sec:jac}, if the automorphism group of a $g$-dimensional abelian variety $X$ violates the Hurwitz bound 
\begin{gather}
    \Aut{X,\cL} \leq 168(g-1),
\end{gather}
then it cannot be the Jacobian of a smooth Riemann surface. By construction the varieties $\cB_\tau[\Frg]$ have automorphism groups containing $\cW[\Frg]$, so it is easy to see that these varieties often violate the bound. Comparing with \cref{tab:coxeter}, we see that $|\cW[\Frg]| > 168(g-1)$ for all choices of $\Frg$ except when $\Frg= A_\ell, B_\ell, C_\ell,D_\ell$ with $\ell<5$ and $\Frg=G_2$. However, as mentioned in \cref{sec:universal}, $\cB_\tau[D_4]$ admits additional automorphisms due to the symmetries of the Coxeter diagram. These symmetries enhance the automorphism group from $\cW[D_4]$ to $\cW[F_4]$, which \emph{does} violate the Hurwitz bound. As such, we can discount the $D_4$ case and conclude that $\cB_\tau[\Frg]$ for $\Frg\notin\{A_\ell,B_\ell,C_3, G_2:\ell\leq 4\}$ do not belong to the open Torelli locus $\cT_g$.

This alone is not enough to determine whether there exists a compatible SW curve as reducible varieties may violate the Hurwitz bound but, nonetheless, be the product of lower dimensional Jacobians. However, the reducibility of the varieties $\cB_\tau$ was studied in \cite{CGR06}, allowing us separate these cases. The results are as follows.
\begin{itemize}
    \item For $A_\ell$ with $\ell>2$, the varieties $\cB_\tau[A_\ell]$ are always irreducible. When $\ell=2$, however, $\cB_{\tau}[A_\ell]$ is reducible if and only if $\tau=-\tfrac{1}{2}(-3+\sqrt{-3})$ modulo $\Gamma_0(3)$. In this case, it is isomorphic to the self-product variety $E_\rho\times E_\rho$ where $\rho=\e^{2\pi\I/3}$. Due to the isomorphism $B_{3\tau}[A_2]\cong B_\tau[G_2]$, this same condition applies to the $G_2$ case.
    \item The varieties for $\Frg=B_n$ are always reducible. Through the Hecke transformation $\tau\mapsto -1/2\tau$, the electric phase of $\Frg=C_n$ is also reducible, but $\cB_\tau[C_n]$ is not.
    \item For $\Frg=D_n$ the varieties $\cB_\tau[\Frg]$ are irreducible except possibly in the cases where $E_{\tau}\cong E_{\tau/2}$ for $n$ even and $E_\tau\cong E_{\tau/4}$ for $n$ odd. Therefore, this occurs when $\tau$ is given by
    \begin{gather}
        \tau = \frac{(aq-d)+\sqrt{(d-aq)^2+4qbc}}{2c},\quad\quad\quad \left[\begin{array}{cc} a & b \\ c & d
    \end{array}\right]\in \SL_2(\bbZ),
    \end{gather}
    with $q=2$ for $n$ even and $q=4$ for $n$ odd. Again, due to the enhanced automorphism groups of the $D_n$ root system, this irreducibility also applies to $\Frg=C_n$ and $\Frg=F_4$.
    \item Finally, the varieties $\cB_\tau[E_n]$ are irreducible for any value of $\tau$ and $n$.
\end{itemize}
With this in mind, we conclude that the varieties $\cB_\tau[\Frg]$ for $\Frg\notin\{A_\ell,B_n, C_3,G_2:\ell\leq 4,n\in\bbN\}$ cannot be the Jacobian of a bouquet of lower dimensional Riemann surfaces in addition to smooth genus $n$ Riemann surfaces, thus ruling out a minimal SW system for these phases.

\subsubsection{Low genus exceptions for \texorpdfstring{$A_\ell$}{A-type} sYM}

In the $A_\ell$ case, we saw that the varieties $\cB_\tau$ are never Jacobian if $\ell>4$. Nevertheless, the Weyl group of $A_\ell$ is still large enough to constrain the possible SW curve for $\ell\leq 4$. Indeed, in \cite{Argyres:2022fwy} the curve for $\ell=2$ was found by analysing the possible genus 2 Riemann surfaces with enhanced automorphism groups. Here we comment on the analogous problem for $\ell=3$ and $\ell=4$.

\medskip
\noindent {\bf Genus 3.} For $A_3$ sYM, we saw that there are two $S$-duality orbits--- one with a single completely reducible fibre and one containing the irreducible variety $\cB_\tau$.  For the former, a bouquet of elliptic curves can be used as a possible SW curve, while the later requires an irreducible genus 3 Riemann surface. 

Genus 3 Riemann surfaces fall into two classes. The first are given by hyperelliptic curves, while the second are that of plane quartics. Each of these classes define a subvariety of the coarse moduli space $\cM_3$ that are stratified by loci corresponding to enhanced automorphism groups (see \cite{MR4574430}, for example). We are therefore looking for 1-dimensional strata whose corresponding surfaces contain $S_4$ in their automorphism group. By comparing with the tables of \cite{MR4574430}, we see that this does not occur in the hyperelliptic case. On the other hand, the plane quartic case contains the Kuribayashi-Sekita family of genus $3$ Riemann surfaces \cite{KS79,Ries95}
\begin{gather}
    \cF_\alpha=\{[x,y,z]\in\bbP^2:x^4 + y^4 +z^4 = \alpha(x^2 y^2+x^2 z^2+ y^2 z^2)\},
\end{gather}
which are invariant under $S_4$. Furthermore, it can be shown that the Jacobian of $\cF_\alpha$ is isomorphic to $E^2\times (E/K)$ with $K\cong \bbZ_4$ \cite{Ries95}, matching the expectation for the irreducible $S$-duality orbit of $A_3$ sYM. We therefore conclude that the Kuribayashi-Sekita family can give a minimal genus SW curve in the automorphism frame picture of \cite{Argyres:2022fwy}.

\medskip
\noindent {\bf Genus 4.} Unlike the genus $3$ case, there does not exist a $1$-dimensional family of varieties invariant under the Weyl group $\cW[A_4]=S_5$. Instead, among genus 4 Riemann surfaces, there is a unique curve with automorphism group $S_5$--- Bring's curve $B$. It can be described by 
\begin{gather}
    B = \left\{[x_1,x_2,x_3,x_4,x_5]\in\bbP^4:\, \sum_{i=1}^{5} x_i^k=0, \ k=1,2,3 \right\}.
\end{gather}
The fact that this does not belong to a 1-dimensional family has the interesting consequence that if we wish to use this as an SW curve for $A_4$ sYM, we can only use it for specific values of the marginal coupling $\tau$. In particular, it is known that the complex modulus of Bring's curve can be put into the form \cite{GAR00,MR4679429}
\begin{gather}
    Z = \tau_0\left[\begin{array}{cccc}
        \ph{-}4 & -1 & -1 & -1 \\
        -1 & \ph{-}4&-1&-1  \\
         -1&-1& \ph{-}4 &-1 \\
         -1&-1&-1& \ph{-}4
    \end{array}\right],
\end{gather}
where $\tau_0$ is determined up to a $\Gamma_0(5)$ transformation by $j(\tau_0)=-(5\times 29^3)/2^5$ and $j(5\tau_0)=-25/2$. From this it can be explicitly checked that the Jacobian is isomorphic to $E_{5\tau_0}^3\times E_{\tau_0}$ as a complex torus. Together these two conditions mean that if $\tau$ is tuned to $5\tau_0$, then Bring's curve reproduces the correct fibre in the magnetic phase. Elsewhere on the conformal manifold, however, there is no minimal genus solution. 

\subsection{Higher genus curves and Prym varieties}

Despite the results of the previous section show that a minimal genus Seiberg-Witten curve is often ruled out by the Hurwitz automorphism theorem, it is known that there exist higher genus solutions \cite{DHoker:1997hut,DHoker:1998xad}. These curves arise as the spectral curves of (twisted) Calogero-Moser systems from which an ACI system can be obtained via the notion of \emph{Donagi-Prym varieties}. Let us show that taking the $\cN=4$ limit of these results recover the ACI systems we have derived, up to a subtlety. 

Given an integrable system, a Lax pair $(L,M)$ is a pair of matrices such that the system of equations defining the system can be put in the form
\begin{gather}
    \frac{\diff L}{\diff t} =[L,M].
\end{gather}
The spectral curve of this system is then given by
\begin{gather}\label{eq:spectral}
    C = \{ (\lambda,z)\in\bbC^2: \det(L(z)-\lambda\,\mathrm{id})=0\},
\end{gather}
where $\lambda$ is the so-called spectral parameter. Often it can be shown that the flow of the integrable system linearises on an appropriate abelian subvariety of $\jac(C)$ \cite{vanhaecke96}. It is precisely this subvariety that forms the fibre of the corresponding ACI system but there is a subtlety in extracting the correct subvariety. Namely, there may be several choices of Lax pairs of differing dimensions that lead to Jacobians of differing dimensions and different abelian subvarieties. As such, there should be, in some sense, a natural choice common to all possible choice of Lax operators. In certain cases, this was answered by Donagi in \cite{donagi93} using a generalisation of Prym varieties as follows. 

Viewing \cref{eq:spectral} as defining a covering $p:C\rightarrow \tilde{C}$ of the curve $\tilde{C}$ parameterised by $z$, we further suppose that this is a tame Galois covering with deck group $G$. The action of the deck group defines an action on $H^0(\Omega_C)$ which, in turn, defines an action on the Jacobian of $C$. For any given irreducible representation of $G$, we can therefore define the corresponding \emph{Donagi-Prym variety} $\mathrm{Prym}_\rho(C)$ to be the connected component of $(\jac(C)\otimes\rho^*)^G$ \cite{MR1836786}. If every irreducible representation of $G$ can be realised over $\bbQ$, then it is known that the Jacobian of $C$ decomposes up to isogeny as \cite{donagi93}
\begin{gather}\label{eq:decomp}
    \jac(C)\simeq\bigoplus_j\rho_j\otimes\mathrm{Prym}_{\rho_j}(C).
\end{gather}
This is the case for Weyl groups, which we now restrict to. As one would expect, this decomposition will differ depending on the representation of $G$ that $L$ is in. However, for special choices of representation, there is a natural common component of the decomposition. In particular, if $H\subsetneq G$ is a proper Weyl subgroup of $G$, then choosing the representation of $L$ to be the permutation representation of $G$ acting on $H$-cosets ensures that the Prym variety corresponding to the reflection representation acting on $\Gamma_r$ appears in the decomposition with positive multiplicity. As such, the appropriate abelian subvariety to choose is one isogenous to $\Gamma_r\otimes \mathrm{Prym}_{\Gamma_r}(C)$.

Returning to sYM, in \cite{DHoker:1997hut,DHoker:1998xad} it was shown that the appropriate integrable systems for $\cN=2^*$ sYM are the (twisted) Calogero-Moser systems. The coupling of the Calogero-Moser potential $m$ is identified with the mass of the adjoint scalar breaking the $\cN=4$ supersymmetry to $\cN=2^*$, so taking the limit $m\rightarrow 0$ provides a spectral curve for $\cN=4$ sYM from that of the Calogero-Moser system. In particular, the spectral curve degenerates into a bouquet of elliptic curves and the corresponding Jacobian becomes $E_\tau^{d}$ where $d$ is dimension of the permutation representation the Lax operator is in. As this is the Jacobian of a reducible curve, the polarisation is given by the principal product polarisation inherited from each $E_\tau$ factor. Writing this variety as $V/\bbZ[\tau]^d$, the appropriate subvariety can be found by decomposing $\mathrm{Re}\,V$ into irreducible $\bbQ[G]$-modules. Amongst this decomposition will be the $\bbQ[G]$-module $\mathrm{Re}\,W$ corresponding to the reflection representation. Assuming this submodule appears with multiplicity 1 for simplicity, then the subvariety of $E_\tau^d$ we require is
\begin{gather}
    W/(W\cap\bbZ[\tau]^d) \cong W/(\Gamma\oplus \tau\Gamma)=\Gamma\otimes E_\tau,
\end{gather}
where $\Gamma$ is a lattice invariant under the reflection representation. As we have seen in previous sections, $\Gamma$ is not necessarily unique, but it will be a refinement of $\Gamma_r$. Therefore, this variety is isogenous to the varieties $\Gamma_r\otimes E_\tau$ considered in \cite{EL76}. Furthermore, the polarisation restricted to these subvarieties are necessarily Weyl invariant and therefore related to the polarisation in \cref{eq:cartan} via isogeny also.

This procedure is easiest to see in the $A_\ell$ case where the lowest dimensional permutation representation is $(\ell+1)$-dimensional. Writing $\jac(C)=V/\bbZ[\tau]^{\ell+1}$, consider the Weyl invariant subspace
\begin{gather}
    W=\left\{(x_1,\ldots,x_{\ell+1}):\sum_{i=1}^{\ell+1} x_i=0\right\}\subset V.
\end{gather}
The intersection $W\cap \bbZ[\tau]^{\ell+1}$ is given by $\Gamma_r\otimes\bbZ[\tau]$ and the principal polarisation restricts to a Weyl invariant polarisation in \cref{eq:cartan} \cite{GAR00}. As such, we conclude that the abelian subvariety $W/(W\cap \bbZ[\tau]^{\ell+1})\cong \Gamma_r\otimes E_\tau$ is precisely the relative variety for $\Frg=A_\ell$ discussed in \cref{sec:relative}.\footnote{To see the isogeny in \cref{eq:decomp}, note that the complementary subvariety is the diagonally embedded $E_\tau\subset E_\tau^{\ell+1}$. This is invariant under the $S_{n+1}$ action and corresponds to the trivial representation in the decomposition. As this is complementary to $\Gamma_r\otimes E_\tau$, there exists an isogeny as in \cref{eq:decomp} \cite[sec. 5]{BL92}.} 

It is interesting to note that the relative fibres for $\Frg$ non-simply laced are not of the form $\Gamma\otimes E_\tau$ since $\Gamma_r\neq \Gamma_r^\vee$. We therefore conclude that in these cases a further isogeny from the subvariety above is required to get the correct relative variety. In fact, for the cases of $B_2$ and $G_2$ considered in \cite{Argyres:2023eij}, the subvariety of $\jac(C)$ gives an absolute form of the theory with split fibre for the former, while it gives a non-principally polarised fibre for the latter. This is despite the fact that $G_2$ has trivial centre and does not admit any 1-form symmetries.

\section{Conclusions and future directions}
In this work we have formulated the refinement/covering map procedures of \cite{Gaiotto:2010be,Argyres:2022kon} in terms of isogenies of abelian varieties and ACI systems, recovering the defect group and Freed-Segal-Moore non-commutativity condition from a purely $4d$ perspective. Using the example of $\cN=4$ sYM theories, we constructed both the relative ACI system and various absolute ACI systems for many examples, testing their validity by probing their $S$-duality group solely using the abelian fibres of the system. We also saw that many phases of such theories do not admit a description via Jacobians of Seiberg-Witten curves of minimal genus, thus providing simple Lagrangian examples of theories that require a different description. 

Additionally, this work opens up several avenues to consider next, some of which we mention here.
\begin{enumerate}
    \item In \cite{Closset:2023pmc}, the authors carry out a systematic study of all rank 1 theories with non-trivial one-form symmetries and understand the SW curves in terms of isogenies. Interestingly, they also show that the Mordell-Weil group of a compactified version of the total space of the (relative) SW system appears to encode the defect group of theory. Furthermore, in moving to an absolute version of the theory, they present examples where the Mordell-Weil group encodes potential non-invertible symemtries of the theory. As the rank 1 SW set up is equivalent to the ACI system, it is interest to consider generalisations of their set up using compactifications of the ACI system instead. It would interesting to see if the Mordell-Weil group could be used at higher ranks to understand more general categorical symmetries.
    \item By performing the isogenies from the relative fibre to an absolute fibre, we are able to extract the period matrix of the latter via knowledge of $\ker f$. In the $B_2$ example, this allowed us to identify the variety as the Jacobian of a curve, in accordance with \cite{Argyres:2023eij} and test its $S$-duality properties. A simple extension of this work would be to attempt to carry out the same procedure in rank 3 where the Schottky problem is understood. If we obtain an irreducible variety, we can compare with the known stratification of the moduli space of genus 3 Riemann surfaces via automorphism groups and identify an SW curve in the automorphism frame picture of \cite{Argyres:2022fwy}. By replicating their work, we can then transform the curve to the usual `canonical frame' and restore dependence on CB parameters to obtain new minimal genus SW solutions.
    \item  Recently, variants of the Calogero-Moser systems were constructed with the aim of reproducing the global forms of absolute $\cN=2^*$ theories \cite{Damia:2025bla}. It would be interesting to understand the effect of their alterations have on the Lax pairs of the theory and study its spectral curves and Prym varieties to see if one can obtain the ACI systems for all global forms using the usual spectral curve analysis.
\end{enumerate}

\noindent {\bf Acknowledgements.} It is a pleasure to thank P. Argyres, M. Del Zotto, S. Meynet and M. Weaver for numerous helpful conversations throughout the preparation of this work. The author is also particularly grateful to the authors of \cite{SKLines} for coordinating the release our works. RM is supported by a Knut and Alice Wallenberg foundation postdoctoral scholarship in mathematics.

\appendix
\section{The moduli space of polarised abelian varieties}
As morphisms of abelian varieties play an important role in the body of this paper, let us briefly review the moduli space of abelian varieties. For a more detailed discussion, we refer the reader to \cite{BL92}. Note, however, that our conventions for period matrices differ, resulting in slightly different actions of $\Sp_{2g}^D(\bbZ)$ on $\bbH_g$.

Given any $g$-dimensional abelian variety $X$ with polarisation $H$, by choosing a symplectic basis for $\mathrm{Im}\,H$, we can always put the associated period matrix into the form $\Pi=[J,Z]$ where $J=\mathrm{diag}(j_1,\ldots,j_g)$ is the type of the polarisation and $Z$ is a symmetric $g\times g$ matrix such that $\mathrm{Im}\, Z$ is positive definite. The map $X\mapsto Z$ therefore associates to each abelian variety of a fixed type $D$ a point in the Seigel upper half space
\begin{gather}
    \bbH_g = \{ M\in \mathrm{Mat}(g\times g,\bbC): M^T=M,\  \mathrm{Im}\, M\  \text{positive definite}\}.
\end{gather}
However, as there could be several symplectic bases for $\mathrm{Im}\,H$, the map $X\mapsto Z$ is not entirely well-defined unless we account for possible isomorphisms of $X$. In order to so, recall that for a homomorphism $f:X\rightarrow Y$ between complex tori, the period matrices $\Pi_X$ and $\Pi_Y$ are related by
\begin{gather}
    \Pi_X = \rho_a(f)^{-1} \Pi_Y \rho_r(f).
\end{gather}
while the polarisations are related by \cref{eq:comm}. If $f$ is an isomorphism, it must also preserve the polarisation in a symplectic basis. Therefore $\rho_r(f)\in\Sp_{2g}^J(\bbZ)$ where
\begin{gather}
    \Sp_{2g}^J(\bbZ) = \{ M\in \GL_{2g}(\bbZ): M^T \Omega M= \Omega\},\quad \Omega=\left[\begin{array}{cc}
        \ph{-}\mathbf{0} & J \\
         -J & \mathbf{0}
    \end{array}\right].
\end{gather}
Writing $\rho_r(f)$ in block form, we can write $\Pi_X$ as
\begin{gather}
    \Pi_X = \rho_a(f)^{-1} [J, Z] \left[\begin{array}{cc}
        A & B \\
        C & D
    \end{array}\right] = \rho_a(f)^{-1}[J A+Z C,J B+Z D].
\end{gather}
In order to ensure $\Pi_X$ has the form $[J,Z']$, we must therefore have $\rho_a(f) = ( J A+Z C) J^{-1}$. This gives us the relation
\begin{gather}\label{eq:symaction}
    Z' =  (J A J^{-1}+ Z C J^{-1})^{-1}(J B+ Z D).
\end{gather}
This provides an action of $\Sp_{2g}^J(\bbZ)$ on $\bbH_g$ that relates isomorphic abelian varieties. As such, the space $\cA_g^J=\bbH_g/\Sp_{2g}^J(\bbZ)$ is the (coarse) moduli space of polarised abelian varieties of type $D$. When the polarisation is principal, we omit the $J$ in this notation.

Given a point $Z\in\bbH_g$, the elements of $\Sp_{2g}^J(\bbZ)$ which fix $Z$ correspond to automorphisms of the associated abelian variety. For a generic $Z$, only $\pm\mathrm{id}$ fixes it, so the corresponding abelian varieties have a $\bbZ_2$ automorphism group. However, for specific values of $Z$, one can obtain a much larger stabiliser inside $\Sp_{2g}^J(\bbZ)$. These points belong to the so-called \emph{singularity locus} of the moduli space and give rise to abelian varieties with enhanced automorphism groups. Equally, one could consider the moduli space as a quotient stack instead, which remains smooth and keeps track of the stabilisers of the group action at each point. The points in the singularity locus then correspond to points isomorphic to the classifying space $BG\cong [*/G]$ with $G$ larger than $\bbZ_2$. For this reason, these points are sometimes called the stacky points of the moduli space.

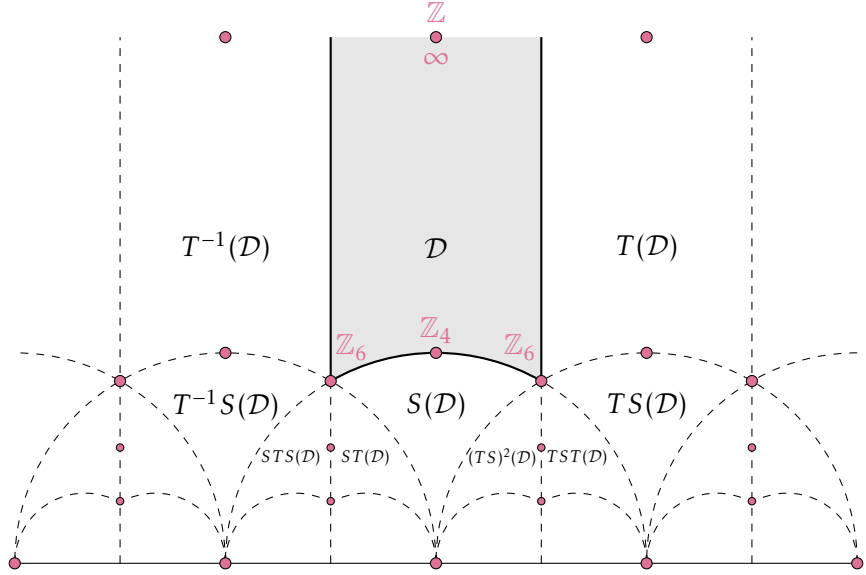
\begin{figure}
    \centering
    \begin{tikzpicture}[scale=1.4]
        \draw (-4,0) -- (4,0);
        \draw[dashed] (0,0) arc (0:180:2);
        \draw[dashed] (2,0) arc (0:180:2);
        \draw[dashed] (-2,0) arc (0:90:2);
        \draw[dashed] (4,0) arc (0:180:2);
        \draw[dashed] (4,2) arc (90:180:2);
        \draw[dashed] (0,0) arc (0:120:2/3);
        \draw[dashed] (-2,0) arc (180:60:2/3);
        \draw[dashed] (2,0) arc (0:120:2/3);
        \draw[dashed] (4,0) arc (0:120:2/3);
        \draw[dashed] (2,0) arc (180:60:2/3);
        \draw[dashed] (-2,0) arc (0:120:2/3);
        \draw[dashed] (-4,0) arc (180:60:2/3);
        \draw[dashed] (0,0) arc (180:60:2/3);
        \draw[dashed] (1,0) -- (1,5);
        \draw[dashed] (3,0) -- (3,5);
        \draw[dashed] (-1,0) -- (-1,5);
        \draw[dashed] (-3,0) -- (-3,5);
        \fill[gray!20] (1,5) -- (1,2*0.866) arc (60:120:2) -- (-1,5) -- (1,5);
        \draw[thick] (1,5) -- (1,2*0.866) arc (60:120:2) -- (-1,5);
        \node at (0,3){$\mathcal{D}$};
        \node at (2,3){$T(\mathcal{D})$};
        \node at (-2,3){$T^{-1}(\mathcal{D})$};
        \node at (0,1.5){$S(\mathcal{D})$};
        \node at (2,1.5){$TS(\mathcal{D})$};
        \node at (-2,1.5){$T^{-1}S(\mathcal{D})$};
        \draw[fill=purple!55] (0,2) circle (1.5pt);
        \draw[fill=purple!55] (1,2*0.866) circle (1.5pt);
        \draw[fill=purple!55] (-1,2*0.866) circle (1.5pt);
        \draw[fill=purple!55] (-2,2) circle (1.5pt);
        \draw[fill=purple!55] (-3,2*0.866) circle (1.5pt);
        \draw[fill=purple!55] (2,2) circle (1.5pt);
        \draw[fill=purple!55] (3,2*0.866) circle (1.5pt);
        \draw[fill=purple!55] (0,5) circle (1.5pt);
        \draw[fill=purple!55] (-2,5) circle (1.5pt);
        \draw[fill=purple!55] (2,5) circle (1.5pt);
        \draw[fill=purple!55] (0,0) circle (1.5pt);
        \draw[fill=purple!55] (-2,0) circle (1.5pt);
        \draw[fill=purple!55] (-4,0) circle (1.5pt);
        \draw[fill=purple!55] (2,0) circle (1.5pt);
        \draw[fill=purple!55] (4,0) circle (1.5pt);
        \draw[fill=purple!55] (-1,1.1) circle (1pt);
        \draw[fill=purple!55] (1,1.1) circle (1pt);
        \draw[fill=purple!55] (3,1.1) circle (1pt);
        \draw[fill=purple!55] (-3,1.1) circle (1pt);
        \draw[fill=purple!55] (-3,0.59) circle (1pt);
        \draw[fill=purple!55] (-1,0.59) circle (1pt);
        \draw[fill=purple!55] (1,0.59) circle (1pt);
        \draw[fill=purple!55] (3,0.59) circle (1pt);
        \node [right] at (-1,1){\scalebox{0.6}{$ST(\cD)$}};
        \node [left] at (-1,1){\scalebox{0.6}{$STS(\cD)$}};
        \node [left] at (1.05,1){\scalebox{0.6}{$(TS)^2(\cD)$}};
        \node [right] at (0.95,1){\scalebox{0.6}{$TST(\cD)$}};
        \node [above] at (0,2){\textcolor{purple!55}{$\bbZ_4$}};
        \node [above] at (0,5.05){\textcolor{purple!55}{$\bbZ$}};
        \node [below] at (0,4.95){\textcolor{purple!55}{$\infty$}};
        \node [above] at (-1+0.2,2*0.866+0.1){\textcolor{purple!55}{$\bbZ_6$}};
        \node [above] at (1-0.2,2*0.866+0.1){\textcolor{purple!55}{$\bbZ_6$}};
    \end{tikzpicture}
    \caption{A fundamental domain $\mathcal{D}$ for the fractional linear action of $\mathrm{SL}_2(\mathbb{Z})$ on the upper half plane. Shown are some images of $\mathcal{D}$ and the stacky points which have are fixed under some subgroup of $\mathrm{SL}_2(\mathbb{Z})$.}
    \label{fig:fund_domain}
\end{figure}

Note that given two principally polarised abelian varieties $A_1$ and $A_2$ of dimensions $g_1$ and $g_2$ respectively, one can define a principally polarised variety of dimension $(g_1+g_2)$ by polarising $A_1\times A_2$ with the product polarisation. In terms of the corresponding ample line bundles $\cL_i$, the polarisation takes the form
\begin{gather}
    \cL = p_1^* \cL_1 \otimes p_2^*\cL_2,
\end{gather}
where $p_i:A_1\times A_2\rightarrow A_i$ is the projection map onto each factor. The gives an embedding of $\cA_{g_1}\times\cA_{g_2}\hookrightarrow \cA_{g_1+g_2}$. Fixing a dimension $g$, we define the decomposable locus of $\cA_g$ to be the union of spaces of the form $\cA_{g_1}\times\cdots\times\cA_{g_n}$ with $g_1+\ldots+g_n=g$ realised by the above embedding. The complement of this, the \emph{indecomposable component} of $\cA_g$, which we denote by $\cA_g^{\mathrm{in}}$, therefore classifies the $g$-dimensional varieties which are not the product of lower dimensional ones.

\begin{example}
In the case of $g=1$, we recover the well-known moduli space of elliptic curves $\bbH_1/\SL_2(\bbZ)$ with the usual fractional linear action. A fundamental domain for this action is drawn in \cref{fig:fund_domain}, where we've noted the stacky points corresponding to enhanced automorphism groups. Within $\cD$ these are given by $\tau=\e^{2\pi \I/6}$ and $\e^{2\pi\I/3}$ with a $\bbZ_6$ automorphism group and $\tau=\I$ with an enhancement to $\bbZ_4$. The well-known $j$-invariant given by
\begin{gather}
j(\tau)=1728\frac{E_4(\tau)^3}{E_4(\tau)^3-E_6(\tau)^2},
\end{gather}
where $E_i(\tau)$ are Eisenstein series, provides a bijection between $\cA_1$ and $\bbC$. The stacky points are often identified by their $j$-invariant value; $j(\tau)=0$ when $\tau=\e^{2\pi \I/3}$ or $\tau=\e^{2\pi\I/6}$ and $j(\tau)=1728$ when $\tau=\I$.
\end{example}

\phantomsection
\addcontentsline{toc}{section}{References}
\bibliography{shw.bib}
\end{document}